\documentclass[11pt,a4paper]{article}

\usepackage{jcappub}
\usepackage[english]{babel}
\usepackage{subcaption}
\usepackage{float}
\usepackage[export]{adjustbox}
\usepackage[normalem]{ulem} 
\usepackage{cancel}

\usepackage{xcolor}
\definecolor{mycolor}{RGB}{255,0,255}

\newcommand{\mr}[1]{\mathrm{#1}} 

\newcommand{\diff}{\mathrm{d}}

\AtBeginDocument{} 
\AtBeginDocument{} 

\AtBeginDocument{} 
\newcommand{\e}{\mr e} 
\newcommand{\I}{\mathrm{i}} 

\usepackage{bm}
\let\vec\bm

\newcommand{\bifrost}{\texttt{BIFROST}}

\newcommand{\aeq}{a_\mathrm{eq}}
\newcommand{\keq}{k_\mathrm{eq}}
\newcommand{\fb}{f_\mathrm{b}}
\newcommand{\fnu}{f_\nu}
\newcommand{\OmegaL}{\Omega_\Lambda}
\newcommand{\OmegaM}{\Omega_\mathrm{m}}
\newcommand{\OmegaC}{\Omega_\mathrm{c}}
\newcommand{\OmegaB}{\Omega_\mathrm{b}}
\newcommand{\OmegaR}{\Omega_\mathrm{r}}

\newcommand{\Msol}{\mathrm{M}_\odot}
\newcommand{\kpc}{\mathrm{kpc}}
\newcommand{\Mpc}{\mathrm{Mpc}}
\newcommand{\kms}{\mathrm{km}\,\mathrm{s}^{-1}}

\graphicspath{{Figures/}}

\title{Structure formation with primordial black holes: collisional dynamics, binaries, and gravitational waves}
\author[a,b]{M. Sten Delos,}
\author[b]{Antti Rantala,}
\author[c,d]{Sam Young,}
\author[b]{Fabian Schmidt}
\affiliation[a]{Carnegie Observatories,\\813 Santa Barbara Street, Pasadena, CA 91101, USA}
\affiliation[b]{Max Planck Institute for Astrophysics,\\Karl-Schwarzschild-Stra{\ss}e 1, 85748 Garching, Germany}
\affiliation[c]{Department of Physics and Astronomy, University of Sussex,\\Brighton BN1 9QH, United Kingdom}
\affiliation[d]{Instituut-Lorentz for Theoretical Physics, Leiden University,\\Niels Bohrweg 2, 2333 CA Leiden, The Netherlands}

\emailAdd{mdelos@carnegiescience.edu}

\emailAdd{anttiran@mpa-garching.mpg.de}

\emailAdd{sam.young@sussex.ac.uk}

\emailAdd{fabians@mpa-garching.mpg.de}

\abstract{
  Primordial black holes (PBHs) could compose the dark matter content of the Universe. We present the first simulations of cosmological structure formation with PBH dark matter that consistently include collisional few-body effects, post-Newtonian orbit corrections, orbital decay due to gravitational wave emission, and black-hole mergers. We carefully construct initial conditions by considering the evolution during radiation domination as well as early-forming binary systems. We identify numerous dynamical effects due to the collisional nature of PBH dark matter, including evolution of the internal structures of PBH halos and the formation of a hot component of PBHs. We also study the properties of the emergent population of PBH binary systems, distinguishing those that form at primordial times from those that form during the nonlinear structure formation process. These results will be crucial to sharpen constraints on the PBH scenario derived from observational constraints on the gravitational wave background. Even under conservative assumptions, the gravitational radiation emitted over the course of the simulation appears to exceed current limits from ground-based experiments, but this depends on the evolution of the gravitational wave spectrum and PBH merger rate toward lower redshifts.
  }

\begin{document}

\maketitle
\flushbottom

\section{Introduction}

Primordial black holes (PBHs) can form in the early Universe if fractional density perturbations on horizon scales are of order unity. While primordial perturbations are only of order $10^{-4}$ on the large scales directly accessible to observations, they could be enhanced on much smaller scales by a variety of mechanisms. Since any peculiar velocities of the PBHs decay rapidly in the early Universe, the PBHs effectively constitute a cold matter (``dust'') component. Thus, if PBHs exist, they could make up a substantial fraction or even all of the dark matter. In fact, PBHs constitute the heaviest ``elementary'' dark matter, and placing constraints on PBHs can be seen as putting an upper limit on the dark matter mass.
We refer readers to refs.~\cite{Carr:2020gox,Escriva:2022duf,Carr:2024nlv} for overviews on the formation and phenomenology of PBHs. 

Gravitational waves arising from mergers of black holes were first detected in September 2015 by the LIGO scientific collaboration \cite{LIGOScientific:2016aoc}. This detection was followed by many further observations by the LIGO, Virgo, and KAGRA collaborations \cite{LIGOScientific:2016sjg,LIGOScientific:2017bnn,LIGOScientific:2017vox,LIGOScientific:2017ycc,LIGOScientific:2018mvr, LIGOScientific:2020aai,LIGOScientific:2020stg}, collectively referred to as LVK. It was quickly suggested by numerous authors that these black holes may have been primordial in origin and that the event rate and other details could be explained by mergers of binary PBH systems that formed inside dark matter halos in a scenario in which the dark matter consists entirely of stellar mass PBHs \cite{Bird:2016dcv,Clesse:2016vqa,Blinnikov:2016bxu}.

In summary, there is strong motivation to study the formation of structure in the Universe, and the detailed distribution of PBH binaries, in the scenario where dark matter is composed entirely of PBHs. While various analytical approaches have been pursued, this is still to a large extent an open problem. While PBHs behave as cold dark matter (CDM) on large scales, there are two differences from particle CDM on smaller scales. First, there are additional small-scale perturbations due to the (approximately) Poisson process of PBH formation \cite{Afshordi:2003zb}, which occurred whenever a horizon-scale patch exceeded a critical collapse threshold \cite{Escriva:2021aeh}. Second, collisional effects become relevant in structure formation due to the fact that all structures are made up of a finite number of black holes. In the case of $\sim 10~\Msol$ PBHs, typical galactic dark matter halos consist of $\sim 10^{11}$ ``particles'', and collisional effects can become important for lower-mass halos.
Previous simulations of PBH cosmologies \cite{Inman:2019wvr,Liu:2022okz,Zhang:2024ytf} neglected collisional effects, which can be a reasonable approximation only if PBHs are a tiny fraction of the dark matter.

In this paper, we present simulations of a cosmological volume containing an abundance of primordial black holes, aiming to capture these two differences to particle dark matter accurately. 
Unlike in other simulations of structure formation from dark matter, here the constituent particles are actually \emph{physical}. Hence, we attempt to solve all dynamical effects such as short-range interactions, binary formation and evolution, and black hole mergers by using the \bifrost\ hierarchical $N$-body code \cite{Rantala2023a,Rantala2024a,Rantala2024b}.
We will compare with two collisionless simulations: one starting from pseudo-particles (phase space elements) initially on a grid, corresponding to the particle dark matter case; and one starting from the same initial conditions as the collisional simulation, but using softening to suppress two-body interactions. This last simulation contains the additional small-scale perturbations from Poisson sampling, but not the collisional effects.

The PBHs in the simulation have masses following a lognormal distribution centered around $10~\Msol$. The lognormal distribution is a relatively generic prediction for PBHs arising from the collapse of large-amplitude density perturbations in the early universe \cite{Gow:2020bzo}.
It has been claimed that there are several observations, including the LVK black holes and correlations in the cosmic X-ray and infrared backgrounds, which could be due to the existence of stellar mass PBHs, as summarized by refs.~\cite{Clesse:2017bsw,Carr:2023tpt}. However, there are also significant constraints on PBHs in this mass range arising from gravitational lensing, accretion onto PBHs and the LVK merger rate; see ref.~\cite{Carr:2020gox} for a summary of constraints on PBHs and ref.~\cite{Mroz:2024mse} for recent limits from lensing. While it may therefore be considered unlikely that PBHs in this mass range could compose the entirety of dark matter, there are significant uncertainties in the observational bounds due to effects such as clustering of PBHs and dynamical interactions of PBH binaries, which can only be addressed using the results of a study such as that presented here.

Using the simulations presented in this paper, we are, for the first time, able to make a comprehensive study of the cumulative effect of PBH interactions throughout the early epochs of structure formation. This includes the evolution of the PBH halos and the substructure inside them. Further, we study the properties of the PBH binary population as it evolves, providing the first fully numerical and resolved study of the PBHs that could give rise to LVK gravitational-wave observations, or be ruled out by them.

The outline of the paper is as follows. In section~\ref{sec:simulations}, we describe the setup of the simulation and the methods used. Section~\ref{sec:structure} explores how cosmic structure forms and evolves in the presence of collisional PBH interactions. Section~\ref{sec:binaries} discusses the formation and evolution of binary PBHs and the gravitational radiation that results from their mergers. Our findings are summarized in section~\ref{sec:summary}.

\section{Simulating primordial black hole dark matter}\label{sec:simulations}

\begin{figure}
  \centering
  \includegraphics[width=0.8\textwidth]{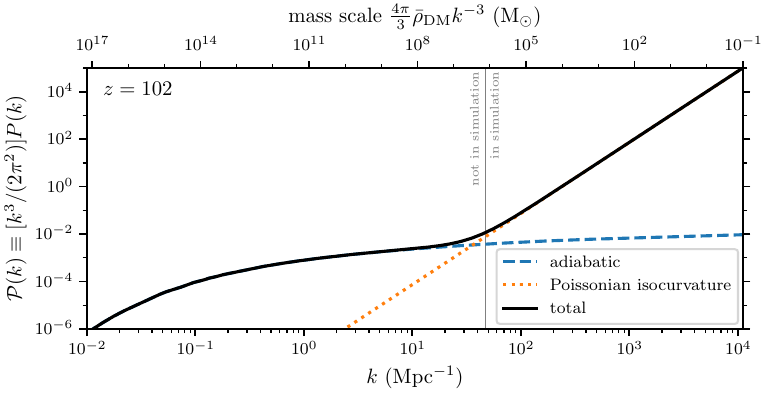}
\caption{Linear dark matter power spectrum for our PBH scenario.
We show the dimensionless form, $\mathcal{P}(k)\equiv [k^3/(2\pi^2)]P(k)$, at $z=102$. The Poissonian spatial distribution of the PBHs manifests itself as an isocurvature white-noise $\mathcal{P}\propto k^3$ contribution to the power spectrum (dotted line), which dominates over the standard adiabatic power (dashed curve) for $k\gtrsim 40~\Mpc^{-1}$.
Scales approximately to the right of the vertical line are represented in the simulation.}
\label{fig:linearpower}
\end{figure}

We adopt a cosmology in which all of the dark matter is in the form of PBHs of average mass $\bar m = 16.487~\Msol$. More precisely, we assume a lognormal mass distribution such that the PBHs have differential number density
\begin{align}\label{PBH_mass-function}
    \frac{\diff n_\mathrm{PBH}}{\diff \ln m}
    =
    \frac{\bar\rho_\mathrm{DM}}{\bar m}
    \frac{\e^{-[\ln(m/m_0)]^2/(2\sigma_{\ln}^2)}}{\sqrt{2\pi}\,\sigma_{\ln}}
\end{align}
per logarithmic interval in the PBH mass $m$, and we set $m_0=10~\Msol$ and $\sigma_{\ln}=1$. Here $\bar\rho_\mathrm{DM}\simeq 33~\Msol\kpc^{-3}$ is the cosmological mean dark matter density.
Forming in causally disconnected patches of the early Universe, the PBHs have an independent random (Poissonian) spatial distribution that gives rise to an isocurvature contribution
\begin{equation}\label{Piso}
    P_\mathrm{iso}(k)=\frac{[D_\mathrm{iso}(a)]^2}{\bar\rho_\mathrm{DM}^2}\int_0^\infty\frac{\diff m}{m}\frac{\diff n_\mathrm{PBH}}{\diff \ln m}m^2
\end{equation}
to the matter power spectrum at the scale factor $a$, where $D_\mathrm{iso}(a)\simeq 2094 a^{0.901}$ is the isocurvature growth function deep in matter domination (see Appendix~\ref{sec:linear}; we define $a=1$ today).
Figure~\ref{fig:linearpower} shows the matter power spectrum associated with this PBH scenario. During the matter era, the Poissonian isocurvature fluctuations start to dominate over the standard adiabatic density variations at wavenumbers $k\gtrsim 40~\Mpc^{-1}$.

We initialize our simulations at $a \simeq 2.9\times 10^{-12}$, which is approximately the scale factor at which $\mathcal{O}(10~\Msol)$ PBHs would form. For our main, ``collisional'', simulation of this PBH scenario, we use the \bifrost\ code that accurately models the $N$-body dynamics of point masses. However, because \bifrost\ does not work in comoving coordinates, it is not well suited for modeling extremely early times during the radiation epoch. Therefore, we use a standard ``collisionless'' cosmological simulation code, \textsc{Gadget-4} \cite{Springel:2020plp}, to advance our PBH simulation up to the scale factor $a=10^{-5}$.
We will also use collisionless simulations to compare with the results of the collisional simulation later.

Our simulations are summarized as follows and detailed in the rest of the section.
\begin{itemize}
    \item A collisional PBH simulation, integrated for $a>10^{-5}$ with \bifrost. For $a<10^{-5}$, we use \textsc{Gadget-4} with severely tightened numerical accuracy parameters and a procedure for recovering PBH binaries.
    \item A collisionless PBH simulation, initialized with the same PBH distribution but integrated at all times using \textsc{Gadget-4} with standard numerical accuracy.
    \item A particle dark matter simulation, initialized with the same realization of the adiabatic density field but without the Poissonian isocurvature fluctuations. Here we also use \textsc{Gadget-4} with standard numerical accuracy.
\end{itemize}
Because \bifrost\ is not cosmological, we cannot use periodic boundary conditions, so we instead simulate a vacuum-bounded spherical portion of a cosmological volume. This is generally a valid approach in simulations of cold dark matter; it only misses the tidal influences of the broader environment and leads to artificial dynamics near the edge of the volume. However, we will see in Section~\ref{sec:structure} that the collisional PBHs do not remain entirely cold, and so as we discuss there, this approach will lead to some numerical artifacts.

\subsection{Initial conditions}\label{sec:collisionless_ics}

To set the initial conditions, we begin with a periodic box of comoving side length 101~kpc.
For the PBH simulations, we fill the box with approximately $128^3$ particles of independently uniformly distributed positions. The particle masses are drawn from the PBH mass distribution.
For comparison, we also initialize a particle dark matter version of these initial conditions with $128^3$ particles initially placed on a grid, such that each simulation particle has mass $16.487~\Msol$.

Next, we include the adiabatic density perturbations that correspond to the cosmological parameters measured by the Planck mission \cite{PlanckAge}.\footnote{
The most widely studied PBH formation mechanism involves modifications to the spectrum of adiabatic perturbations at small scales, and these perturbations can separately influence structure formation \cite{StenDelos:2022jld,Delos:2023fpm}. We neglect this effect, which would require fixing a model for the primordial power spectrum. PBHs around $10~\Msol$ are associated with large-amplitude primordial perturbations at wavenumbers $k\sim 10^6~\Mpc^{-1}$, and although there are limits on how abruptly the power spectrum could return to the scale-invariant level observed at much larger scales \cite{Byrnes:2018txb,Cole:2022xqc}, specific models (e.g.~\cite{Heydari:2021qsr,Franciolini:2022tfm,Heydari:2023xts}) can produce $\mathcal{O}(10~\Msol)$ PBHs while restoring the scale-invariant level of primordial perturbations at wavenumbers below about $10^{3}$ to $10^4~\Mpc^{-1}$, which correspond to typical initial inter-PBH separations in our simulation.
} We use the \textsc{CLASS} Boltzmann code \cite{CLASS} to evaluate the dark matter power spectrum at $z=31$, and then we extrapolate it back to the initial time using the small-scale analytic linear growth function derived by ref.~\cite{Hu:1995en}.
We use the Zel'dovich approximation to set the initial particle displacements and velocities. The same adiabatic perturbations are added to the PBH and particle dark matter simulations. Appendix~\ref{sec:simulation-backgrounds} further details this process.

Finally, we cut out a sphere of radius 21.1 kpc to comprise our simulation region.
This sphere is surrounded by vacuum, a necessity because the collisional code does not employ periodic boundary conditions.
It contains about $80\,000$ PBHs and is close to the mean density, being overdense by only $\delta\equiv (\rho-\bar\rho_\mathrm{DM})/\bar\rho_\mathrm{DM}=0.005$.
However, the region is chosen so that the central $10\,000$ PBHs, occupying a sphere of half the total radius, are overdense by $\delta=0.05$. This overdensity is entirely due to Poisson noise arising from the PBH spatial and mass distribution, as this noise is dominant at these scales.
It corresponds to a 3$\sigma$ fluctuation in a $1.6\times 10^5~\Msol$ mass of dark matter, so about one such fluctuation is expected per $10^{8}~\Msol$ of matter.
We also cut the same spherical region out of the particle dark matter grid, and it has $\delta=0.002$ in the total volume and $\delta=-0.006$ in the half-radius sphere.

\subsection{Collisionless simulations}\label{sec:collisionless_simulations}

To carry out the collisionless simulations, we use a version of \textsc{Gadget-4} that we modified to include the effect of radiation. We also modified it to treat the baryonic component of the matter as a homogeneous background. Although the clustering of baryons can be marginally relevant at the $\mathcal{O}(10~\kpc)$ scales that we study \cite{Bertschinger:2006nq}, we neglect this effect as our focus is on comparing PBH and particle dark matter. 
To confirm the accuracy of these code modifications, we carried out simulations of both the PBH and the particle dark matter periodic boxes (not spheres) and verified that the results match the predictions from linear-order cosmological perturbation theory.
Note that this is true even though the boxes are superhorizon at the initial scale factor $a \simeq 2.9\times 10^{-12}$; one can choose a gauge in which the Newtonian gravitational evolution is accurate at superhorizon scales as long as perturbations are in the linear regime (e.g.~\cite{Fidler:2017pnb}).
The relevant linear theory, code modifications, and validation results are detailed further in appendix~\ref{sec:simulation-backgrounds}.

Using the vacuum-bounded spherical regions, we now execute the particle dark matter grid simulation and the collisionless PBH simulation with our modified version of \textsc{Gadget-4}.
Additionally, we run a third simulation to prepare initial conditions for the collisional PBH simulation. For this purpose, we use greatly tightened numerical accuracy parameters, and we only execute the simulation up to the scale factor $a=10^{-5}$.
In this simulation, forces are softened for PBHs closer than 1.7 comoving parsec (compared to 67~pc in the collisionless simulations), but whenever a PBH pair is found to be closer than 3.4~pc, we record the relative positions and velocities of the PBHs and reconstruct the associated Keplerian binary orbit at $a=10^{-5}$ in the initial conditions for the collisional simulation. The numerical accuracy parameters and the details of this binary restoration procedure are in appendix~\ref{sec:simulation-binaries}, as are tests of its efficacy.

\subsection{Collisional PBH simulation}\label{sec:collisional}

The gravitational collisional dynamics of PBHs can be examined using numerical tools developed over the past six decades for studying the dynamics of star clusters \cite{Aarseth2003}, such as fourth-order (star cluster scale) and regularized (binary system scale) integration techniques. For the collisional PBH simulation we use a modified version of the \bifrost{} code \cite{Rantala2023a,Rantala2024a,Rantala2024b}. \bifrost{} is a novel, fast, and accurate GPU-accelerated direct-summation $N$-body code based on the hierarchical \cite{Rantala2021} fourth-order forward symplectic integrator \cite{Chin1997,Chin2005,Chin2007,Dehnen2011,Dehnen2017}. The fourth-order integrator is responsible for integrating the orbital dynamics of PBHs on spatial scales larger than about $10^{-3}$ pc. In addition to the forward integrator, \bifrost{} includes specialized secular and regularized integration techniques for treating close particle encounters, binaries, triple systems, and small clusters around massive black holes at small scales below about $10^{-3}$ pc.

Weakly perturbed binary PBH systems are treated in a secular manner using a Kepler solver \cite{Mikkola2020} with external Newtonian perturbations from the rest of the simulation. The Peters-Mathews equations of motion for the binary semi-major axis and eccentricity at post-Newtonian order PN2.5 \cite{Peters1963,Peters1964} are responsible for the circularization and shrinking of isolated PBH binary orbits due to gravitational wave radiation reaction forces. The included PN1.0 periapsis advance term is important for quenching von Zeipel-Lidov-Kozai oscillations \cite{vonZeipel1910,Lidov1962,Kozai1962,Ito2019} in hierarchical triple systems and avoiding artificially enhanced (purely Newtonian) PBH merger rates in such systems (e.g. \cite{Mannerkoski2021}). 
Close ($<10^{-3}$ pc) systems with more than two bodies are integrated using an implementation of the efficient and accurate algorithmically regularized integrator LogH \cite{Mikkola1999,Mikkola2002,Mikkola2006,Mikkola2008,Rantala2020}. The equations of motion of the PBHs are post-Newtonian up to the order PN3.5 from refs.~\cite{Thorne1985,Blanchet2014} in the regularized regions and PN1.0 in the rest of the simulation domain, enabling relativistic libration and precession as well as radiation-reaction effects responsible for radiative energy and angular momentum losses. Overall, our PN implementation is very similar to that of ref.~\cite{Mannerkoski2023}.

Two PBHs of masses $m_1$ and $m_2$ are merged when their mutual separation becomes smaller than $12 G (m_1+m_2) / c^2$ or when their merger time scale predicted by the Peters-Mathews formulas becomes shorter than their time step in the code. At the moment of merger, the mass, the spin and the relativistic gravitational-wave recoil kick velocity of the remnant are estimated by using fitting formulas from numerical relativity simulations \cite{Zlochower2015}.
Note that the PBHs in our simulation have zero initial spin, in accordance with standard predictions \cite{DeLuca:2019buf}, and only two PBHs experience a second-generation merger with nonzero spins.

\bifrost{} operates in Newtonian, physical coordinates. The cosmological background is taken into account using a time-dependent external potential leading to an additional acceleration term $\vec{f} = \vec{f}(\vec{r},t)$ in the equations of motion of the particles in the simulation which can be implemented into \bifrost{} in a straightforward manner. The expression for the background force is discussed in appendix~\ref{section:physical-coordinates}.

We run the collisional simulation starting from $a=10^{-5}$ ($z\simeq 10^5$, $t\simeq 7.64\times10^{-5}$~Myr) until $z\sim5.5$ (corresponding to $\sim1$ Gyr) on $1$--$2$ nodes of the \textsc{Freya} cluster hosted by the Max Planck Computing and Data Facility in Garching, Germany. 
We use the following \bifrost{} code user-given accuracy parameters. The forward integrator time-step factors $\eta_\mathrm{ff}$ (free-fall), $\eta_\mathrm{fb}$ (fly-by) and $\eta_\mathrm{\nabla}$ (gradient) are all set to $0.1$. The subsystem neighbor radius defining the sizes of the regularized integration regions is $r_\mathrm{ngb} = 10^{-3}$ pc. The tolerance parameters of the regularized integrator LogH are set to $\eta_\mathrm{GBS} = 10^{-7}$ and $\eta_\mathrm{t}=10^{-2}$. The forward integration interval duration corresponding to the maximum time-step allowed in the code is $\Delta t=10^{-5}$ Myr before $t=1$ Myr, after which we increase the integration interval duration to $\Delta t=10^{-3}$ Myr.

\section{Structure formation with primordial black hole dark matter}\label{sec:structure}

\begin{figure}
  \centering
  \includegraphics[width=.998\textwidth]{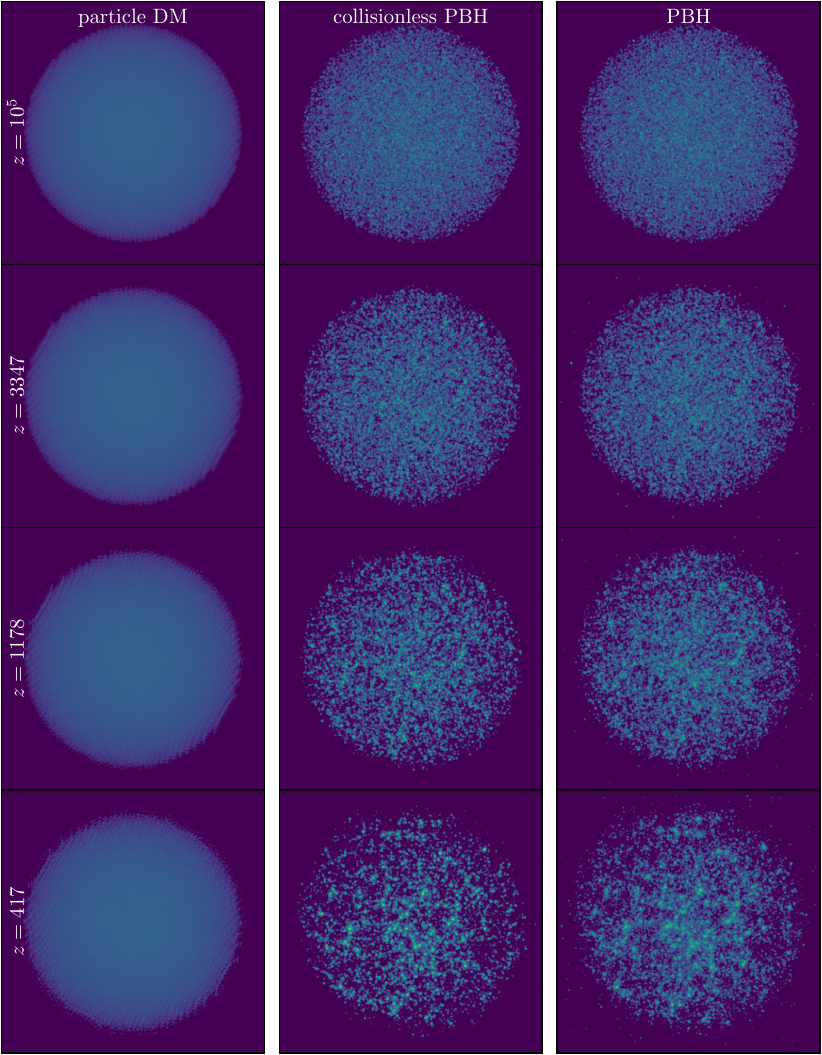}
\caption{Early evolution (from top to bottom) of the simulation volumes. We show the projected density with a logarithmic color scale; lighter is denser. Due to their initial Poisson clustering, the PBH volumes fragment into halos long before any structure is visible in the particle dark matter volume (left). However, the collisionless PBH simulation (middle) artificially forms too much structure compared to the collisional simulation (right).}
\label{fig:fields1}
\end{figure}

\begin{figure}
  \centering
  \includegraphics[width=.998\textwidth]{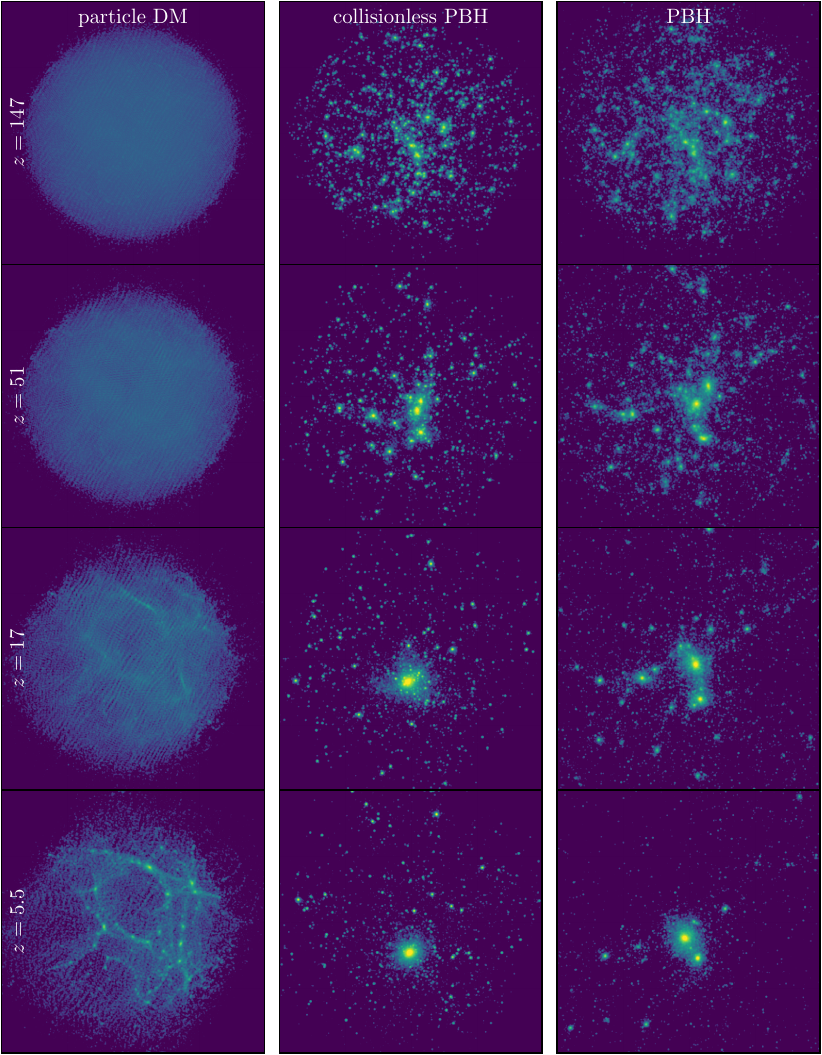}
\caption{Late evolution (from top to bottom) of the simulation volumes, similar to figure~\ref{fig:fields1}. By these redshifts, structures finally become visible in the particle dark matter simulation (left). Both PBH volumes form a halo of mass $10^5$ to $10^6~\Msol$ (see figure~\ref{fig:halo_comparison}), but the abundance of much smaller halos is greatly suppressed by collisional dynamics in the right-hand panels, compared to the collisionless PBH simulation (middle).}
\label{fig:fields2}
\end{figure}

We show the overall evolution of the simulation volumes in figures \ref{fig:fields1} and~\ref{fig:fields2}, comparing the particle dark matter, collisionless PBH, and collisional PBH simulations. The Poissonian white noise in the PBH initial conditions causes the PBH volumes to fragment into clusters of PBHs
already by $z\sim 1000$.
In contrast, the particle dark matter volume does not form significant structure until $z\sim 10$.
For the particle dark matter, some artifacts of the initial grid are visible (i.e., the diagonal patterns).

\subsection{Halo mass functions and substructure}\label{sec:massfunction}

\begin{figure}
  \centering
  \includegraphics[width=\textwidth]{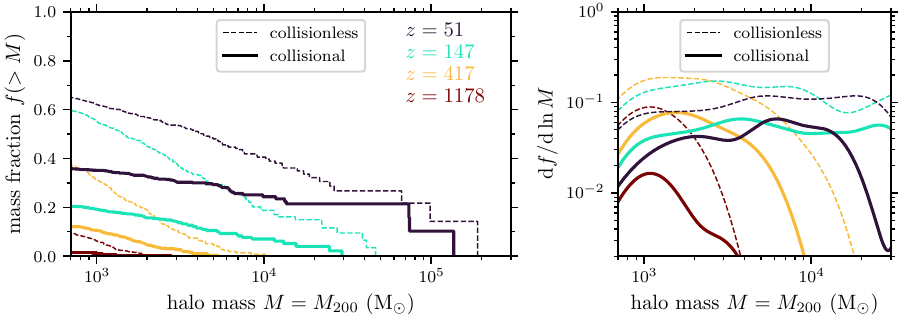}
  \caption{
    Comparison of halo mass functions between the collisional PBH simulation (solid curves) and the collisionless version (dashed curves) at a range of times (different colors). The left-hand panel shows the cumulative mass function. As a function of $M$, we plot the fraction of PBH mass that resides in halos of at least the mass $M$. The right-hand panel shows the differential mass function, i.e., the fraction of mass in halos of mass $M$ per logarithmic mass interval. Here we use a Gaussian kernel of width 0.3 $\e$-folds.
    Collisional dynamics appear to suppress the abundance of low-mass halos by a factor of a few.
    The growth of the most massive halos is also delayed by collisional heating, which we explore in section~\ref{sec:heating}.
}
\label{fig:massfunction}
\end{figure}

Figure~\ref{fig:massfunction} shows the halo mass functions in the PBH simulations.
We identify these halos using the friends-of-friends algorithm \cite{Davis:1985rj} (as implemented in \textsc{Gadget-4}) with the linking length $0.2 \bar n^{-3}$, where $\bar n$ is the cosmological mean number density of PBHs.
Only halos of 32 or more PBHs are considered.
We evaluate the $M_{200}$ masses of these halos, i.e., the masses of the spheres enclosing average mass density 200 times the cosmologically averaged density in PBHs.
For the collisionless PBH simulation, these spherical-overdensity masses are evaluated by centering on the most gravitationally bound particle, as is natively implemented in \textsc{Gadget-4}.
For the collisional simulation, however, the minimum of the potential can be entirely disconnected from the center of the halo, since it can indicate a binary system or close encounter instead. Therefore, in this case we simply use the center of mass of the friends-of-friends group.

The mass functions in figure~\ref{fig:massfunction} quantify the halo abundance trends that were visible in the density fields. Halos of at least $32$ PBHs begin to form around $z\sim 1000$ and comprise more than 10 percent of the mass by $z\sim 400$. We show the mass function only down to redshift $z=51$ because at later times, the mass is dominated by a few large halos. Compared to the (artificial) collisionless PBH simulation, collisional dynamics suppress the abundance of halos below $\sim 3\times 10^4~\Msol$, i.e. halos of fewer than about 2000 PBHs, by a factor of 2 to 3. This suppression of the halo population is much more serious than what standard star cluster evaporation arguments would predict (e.g.~\cite{Jedamzik:2020ypm}).
One possible explanation is that unlike star clusters, PBH halos grow gradually over cosmological time scales. Collisional evaporation is more efficient for clusters of fewer PBHs, so it could have been important during the early stages of halo growth, even for halos that would later become too large to evaporate significantly. Another possible explanation is that many-body PBH interactions can be important, as we will see in section~\ref{sec:heating}.

\begin{figure}
  \centering
  \includegraphics[width=\textwidth]{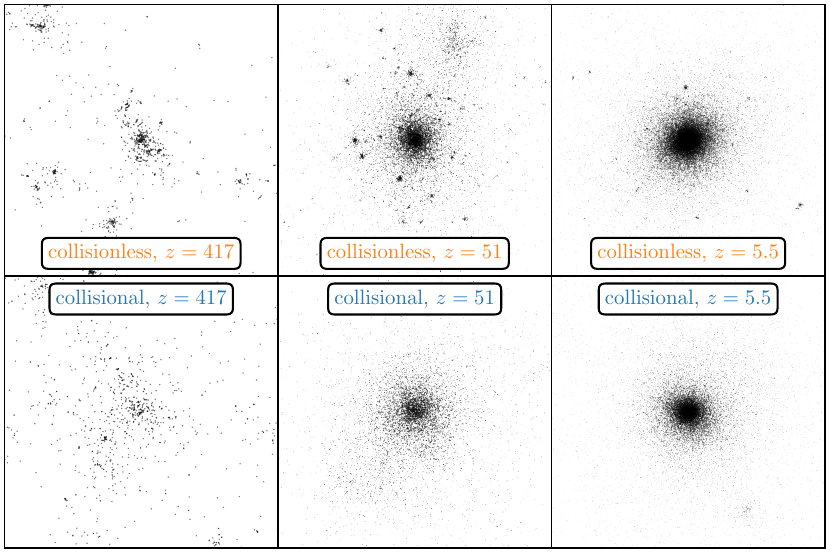}
\caption{
    PBH distribution in a 5.3 comoving kpc box centered on the largest halo. We compare the collisionless (top) to collisional (bottom) simulations at several times (from left to right). Each point represents a PBH; their drawn sizes depend on the time shown, but are the same between simulations. Collisional dynamics represented in the lower panels lead to a lower density, much lower substructure abundance, and a halo of lower mass (although this effect is partially artificial, as we discuss in the text).
    }
\label{fig:subfields}
\end{figure}

In figure~\ref{fig:subfields}, we show a zoomed-in picture of the largest halo. Further trends are visible here. Compared to the collisionless simulation, the PBH halo in the collisional simulation is more diffuse. While the halo in the collisionless simulation has a significant subhalo population, its counterpart in the collisional simulation almost completely lacks substructure.
The absence of substructure in the collisional halo agrees with the results of $N$-body simulations of star cluster assembly; in that context, almost all cluster substructure is erased on a timescale of order $10$ Myr \cite{Rantala2024b}.
Finally, the collisional dynamics appear to delay the growth of the halo, resulting in a significantly less massive system at each time
(an effect already visible in the mass functions in figure~\ref{fig:massfunction}; see also figure~\ref{fig:halo_comparison}). However, the last effect is partially artificial, as we demonstrate next.

\begin{figure}
  \centering
  \includegraphics[width=\textwidth]{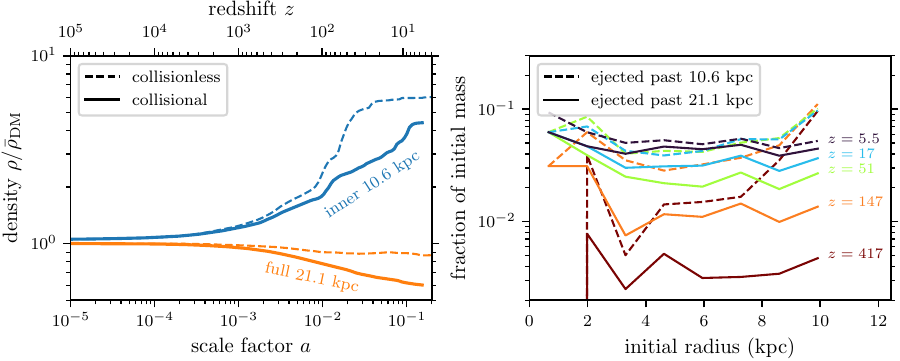}
  \caption{
    Evolution of the average density due to collisional ejection of PBHs from the simulation volume.
    The left-hand panel shows the average density within the whole comoving simulation sphere (orange) as well as that within the central eighth of the volume (blue), 
    both in units of the cosmological mean PBH mass density, $\bar\rho_\mathrm{DM}$.
    Comparing the collisional PBH simulation (solid curves) to the collisionless version (dashed curves), the density in the collisional simulation is consistently lower.
    The right-hand panel illustrates how this effect arises. As a function of the initial position, we show for a range of times (different colors) the fraction of the mass that is ejected from the full volume (solid lines) or from the central portion (dashed lines). Ejected PBHs are distributed almost uniformly in space, aside from an excess in the dashed lines near 10~kpc that corresponds to mass simply drifting out of the volume.
    }
\label{fig:meandensity}
\end{figure}

The left-hand panel of figure~\ref{fig:meandensity} shows how the overall density of the simulation volume evolves. Due to the selection of the initial conditions, the density within the inner 10.6~kpc in comoving radius (half that of the full volume) rises significantly over the course of the simulations. However, the rise is significantly slower in the collisional simulation compared to the collisionless version. Moreover, we also show the density in the full 21.1~kpc radius. This density drops only slightly in the collisionless simulation, as particles leak out of the edges. However, it drops to about 60 percent of the cosmological mean by the end of the collisional simulation, implying that fully 40 percent of the mass in PBHs has been lost from the volume.
The decrease in mass within the full simulation volume is largely an artificial consequence of our use of an isolated vacuum-bounded cosmological volume, since most of it should be replenished in principle by mass lost from neighboring volumes.
The associated density decrease delays structure formation, which explains some of the delayed halo growth visible in figure~\ref{fig:subfields}. For this reason, we expect that the delay would be lessened in a simulation of a larger cosmological volume (or a periodic volume). 
However, the mass loss from the inner 10.6~kpc is not all artificial, since an overdense region should in general lose more mass than is supplied to it by less dense neighboring volumes. Thus, some delay of structure formation within this region is expected to be a real consequence of collisional dynamics.

\subsection{Collisional heating}\label{sec:heating}

Although the decrease of mass within the simulation volume is an artificial consequence of its finite extent, the mechanism for the mass loss is real and may have important consequences for PBH cosmologies.
For a range of times, the right-hand panel of figure~\ref{fig:meandensity} shows the fraction of mass initially at each radius that was lost from the simulation volume. We separately consider mass lost from the full 21.1~kpc sphere and mass lost from the 10.6~kpc central part.
The key feature of this picture is that the mass is lost almost equally from every radius within the simulation volume. This means that it does not primarily come from PBHs with small peculiar velocities leaking out of the edges. Rather, it arises due to a population of PBHs that have been dramatically heated by collisional dynamics.

\begin{figure}
  \includegraphics[width=\textwidth]{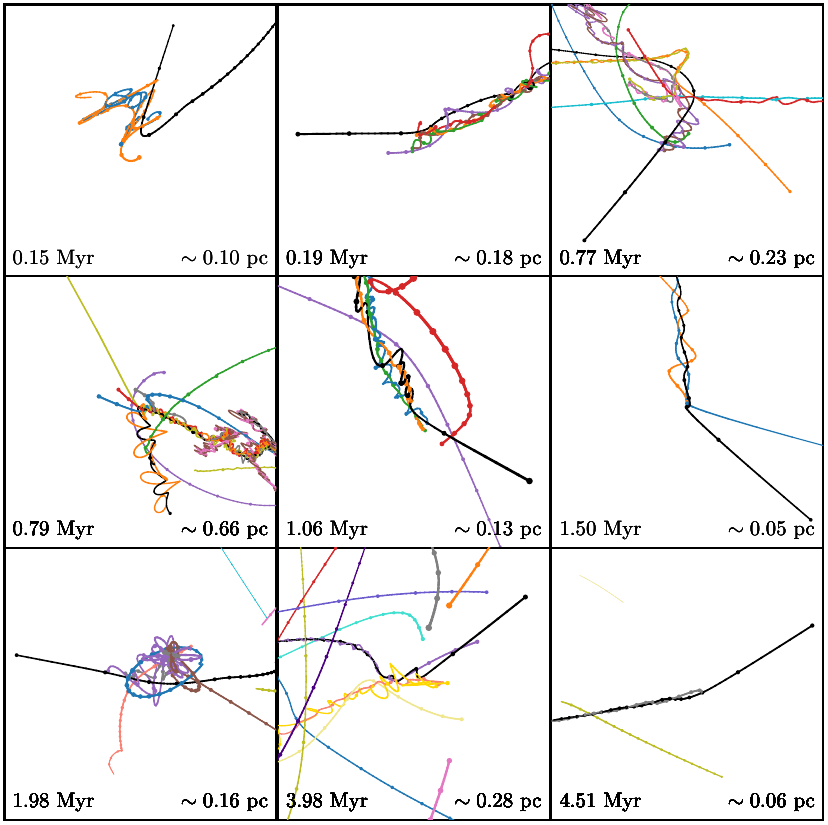}
\caption{
    Representative sample of the dynamics that lead to ejection of PBHs from the simulation volume (see figure~\ref{fig:meandensity}). In each panel, the black PBH is selected arbitrarily from those that were initially within the comoving radius 10.6~kpc but escaped beyond 21.1~kpc by $z=99$. We depict approximately the time and location of the event that resulted in its ejection. In each case, we show the three-dimensional PBH trajectory (in comoving coordinates, interpolated across each 0.01 Myr between simulation snapshots) over the time range that is indicated, and we include other PBHs (different colors) only if they were at any time the nearest or second-nearest to the subject (black) PBH. Circular markers are evenly spaced in time at intervals of $1/20$ the full time range, and the direction of motion is indicated by the marker's presence at the ending time and absence at the starting time. We indicate on each panel the duration and approximate proper length scale depicted (although the length scale does not have a precise definition). Line and marker sizes scale appropriately with distance from the viewer. It is clear that ejection results from many-body (as opposed to two-body) dynamics, except in one case (lower right), where it results from an asymmetrical PBH merger. Often, other PBHs are also ejected.
    }
\label{fig:trajectories}
\end{figure}

Figure~\ref{fig:trajectories} shows a representative sample of these collisional dynamics. Each panel shows an event that led to the loss from the full simulation volume of a PBH (in black) that was initially within the inner 10.6~kpc. It is noteworthy that none of these events can be described as a simple two-body interaction. There is one example in which the heating resulted from a merger between two PBHs due to a gravitational-wave recoil kick. In all other cases, heating of the subject (black) PBH resulted from an interaction between three or more PBHs.

The necessity of many-body interactions to facilitate heating is expected on kinematic grounds. In a two-body collision, PBHs can only exchange their kinetic energy within the host halo, which tends to remain comparable to the halo's potential depth (due to the virial theorem). Consequently, while two-body collisions can eject PBHs from a halo, they cannot eject them at velocities greatly exceeding the escape velocity. On the other hand, with three or more PBHs, it is possible to convert the gravitational potential energy between two PBHs into kinetic energy of a third. Since PBHs can become arbitrarily tightly bound, this enables the heating of PBHs to arbitrarily high speeds. The same is true of PBH mergers, since in this case a portion of the mass of one PBH can be converted into kinetic energy of another (with momentum conserved via anisotropic gravitational radiation).

\begin{figure}
  \centering
  \includegraphics[width=\textwidth]{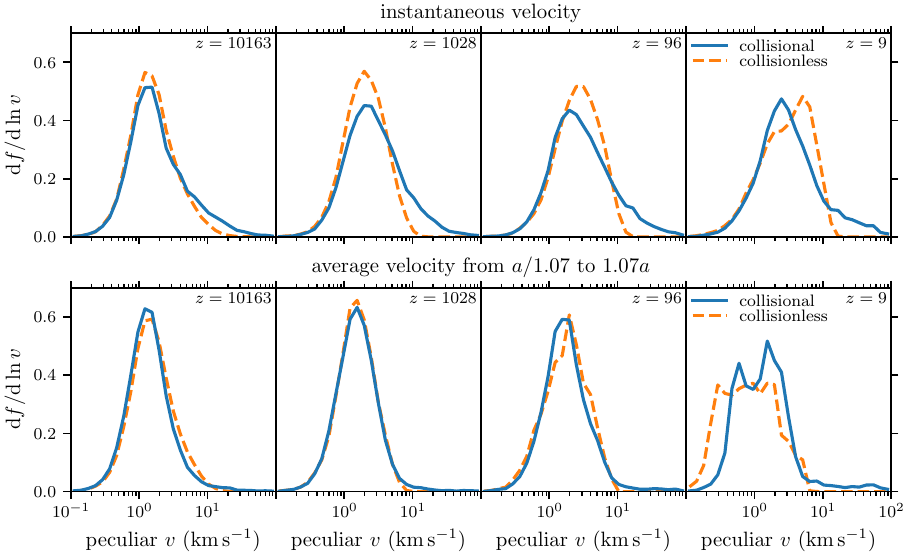}
\caption{
Distribution of peculiar velocities (weighted by PBH mass), comparing the collisional PBH simulation (solid curves) to the collisionless version (dashed curves).
The upper panels show the instantaneous velocities, while the lower panels remove the contribution from short-period orbits by considering the average velocity over about 15 percent of a Hubble time.
From left to right, each panel corresponds to a later time. Evidently, collisional dynamics significantly alter the overall velocity distribution, creating a very-high-velocity tail.
At early times, this tail is mostly associated with orbital motion within hard binaries. However, at late times, the lower right panels indicate that a significant portion of the tail is associated with long-distance streaming.
}
\label{fig:vdist}
\end{figure}

The upper panel of figure~\ref{fig:vdist} shows the distribution of peculiar velocities in the PBH simulations. Collisional dynamics evidently lead to the formation of a long tail in the distribution corresponding to a population of PBHs at very high velocities. Note that not all of these PBHs are necessarily hot in a sense that can affect structure formation. For example, members of a tightly bound PBH binary can have extremely high peculiar velocities even as the peculiar velocity of the binary system remains low.
However, all of them have at least the capacity to heat other PBHs to comparable velocities by interacting with them.
We also point out that these high PBH velocities would suppress the rate at which they accrete gas \cite{Hutsi:2019hlw}, which could impact limits on PBH abundance that derive from radiation emitted during gas accretion \cite{Ziparo:2022fnc}.

To identify the PBHs that are ``hot dark matter'', instead of simply being in tight binaries, we also consider how far each PBH travels over about 15 percent of a Hubble time. This is approximately the dynamical time scale of virialized halos ($\mathcal{O}(100)$ times overdense), so it is far longer than the orbital period of any but the most weakly bound binary pairs. For a scale factor $a$, we identify each particle's comoving displacement $s$ from $a_1\simeq a/1.07$ to $a_2\simeq 1.07a$. For a particle with peculiar velocity $v(a)$ drifting in the absence of peculiar potentials, the displacement $s$ is given by
\begin{align}
    s=\frac{2a v(a)}{H_0\sqrt{\OmegaR}}\left(\mathrm{arsinh}\sqrt{\frac{\aeq}{a_1}}-\mathrm{arsinh}\sqrt{\frac{\aeq}{a_2}}\right),
\end{align}
where $H_0$ is the Hubble constant, $\aeq$ is the scale factor of matter-radiation equality, and $\OmegaR=\OmegaM\aeq$ is the radiation density parameter.
This expression neglects dark energy, which is appropriate since our simulation stops before dark energy becomes significant.
We invert it to find $v(a)$ from $s$; this should be regarded as our definition of the average velocity. The lower panel of figure~\ref{fig:vdist} shows the distribution of these average velocities $v$.
The high-velocity tail is still present at late times, and at $z=9$ (lower right panel), it is nearly log-uniform up to $\sim 100~\kms$.

\subsection{Backreaction onto large-scale dynamics}
\label{sec:backreaction}

The presence of such a hot population of PBHs can affect structure formation at scales much larger than what we are able to simulate.
Hot dark matter suppresses structure in the perturbative regime because it streams out of overdense (or underdense) regions. With less gravitating mass remaining in the density perturbations, they grow in amplitude more slowly. Particles of peculiar velocity $v$ suppress perturbation growth on mass scales smaller than about the Jeans mass, $M\simeq 2.92 v^3 G^{-3/2} \bar\rho_\mathrm{m}^{-1/2}a^{3/2}$ (e.g.~\cite{2008gady.book.....B}), corresponding to that velocity. Here $\bar\rho_\mathrm{m}\simeq 39.5~\Msol\,\kpc^{-3}$ is the present-day cosmological mean matter density, so $\bar\rho_\mathrm{m}a^{-3}$ is the mean matter density at scale factor $a$. Consequently, the streaming of particles with peculiar velocities $v$ exceeding
\begin{align}\label{streaming}
    v \gtrsim 0.7 G^{1/2} M^{1/3} \bar\rho_\mathrm{m}^{1/6}a^{-1/2} && \text{(streaming)}
\end{align}
suppresses structure growth on the mass scale $M$.
Additionally, hot PBHs would escape from (or not accrete onto) virialized structures. The escape velocity from a system of mass $M$ and radius $R$ is around $v=\sqrt{2GM/R}$, but for a virialized halo, $M=\frac{4\pi}{3}R^3\Delta_\mathrm{vir}\bar\rho_\mathrm{m}a^{-3}$, where $\Delta_\mathrm{vir}\simeq 200$ is the virial overdensity. Consequently, particles of peculiar velocity $v$ exceeding
\begin{align}\label{escape}
    v \gtrsim
    4.3 G^{1/2}M^{1/3}\bar\rho_\mathrm{m}^{1/6}a^{-1/2} && \text{(escape)}
\end{align}
would escape from halos of mass $M$.

\begin{figure}
  \centering
  \includegraphics[width=\textwidth]{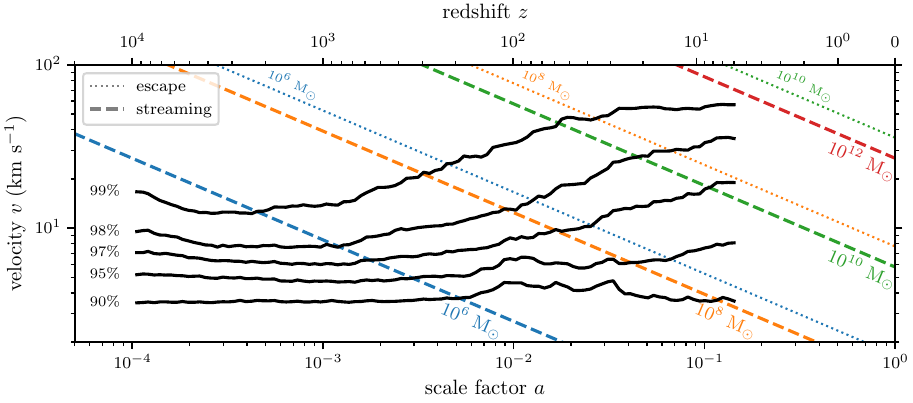}
\caption{
Mass scales that the component of hot PBHs could potentially affect.
The black curves show, as a function of time, the upper percentiles of the distribution of time-averaged PBH velocities (as in the lower panels of figure~\ref{fig:vdist}).
The colored dashed lines mark the velocity of a particle that streams over a distance corresponding to the indicated mass scale $M$, so PBH velocities above these lines would suppress the growth rate of structure on mass scales $\lesssim M$.
The colored dotted lines mark the velocity needed to escape a virialized object of the indicated mass $M$, so PBH velocities above these lines would suppress the masses of halos of mass $\lesssim M$.
This picture suggests that the component of collisionally heated PBHs could have a non-negligible effect on structure at galaxy scales.
Note that the total mass in our PBH simulation is only about $10^6~\Msol$, so a much larger simulation would be needed to include these effects.
}
\label{fig:backreaction}
\end{figure}

In figure~\ref{fig:backreaction}, we explore how the velocity distribution in our PBH simulation compares to the streaming and escape velocity thresholds given by equations (\ref{streaming}) and~(\ref{escape}). We plot the upper percentiles of the distribution of time-averaged peculiar velocities; as discussed in section~\ref{sec:heating}, these are true streaming velocities, because they do not include high-frequency orbital motion.
Comparison with the streaming velocity threshold (dashed lines) suggests that the hot PBH component suppresses the growth rate of structure at the few-percent level on mass scales as high as $\sim 10^{10}~\Msol$ (green dashed lines) and even potentially at the percent level on mass scales exceeding $10^{12}~\Msol$ (corresponding to comoving length scales greater than a few Mpc).
Comparison with the escape velocity threshold (dotted lines) suggests that the hot component could separately suppress the present-day masses of halos up to $\sim 10^{10}~\Msol$ by a few percent.
These effects correspond to a backreaction from small-scale nonlinear structure onto much larger scales, including onto scales on which the evolution would otherwise still be described by linear theory.\footnote{Note that these effects are not captured by the usual formulations of the effective field theory of structure formation \cite{Baumann:2010tm}, since these formulations assume that dark matter remains dynamically cold.} Even though the PBHs that we consider are not much heavier than stars, our analysis suggests that collisional heating of these PBHs is likely to influence the properties and distribution of galaxies at the few-percent level, although more study is needed to quantify the effect precisely.

\begin{figure}
\centering
  \includegraphics[width=0.9\textwidth]{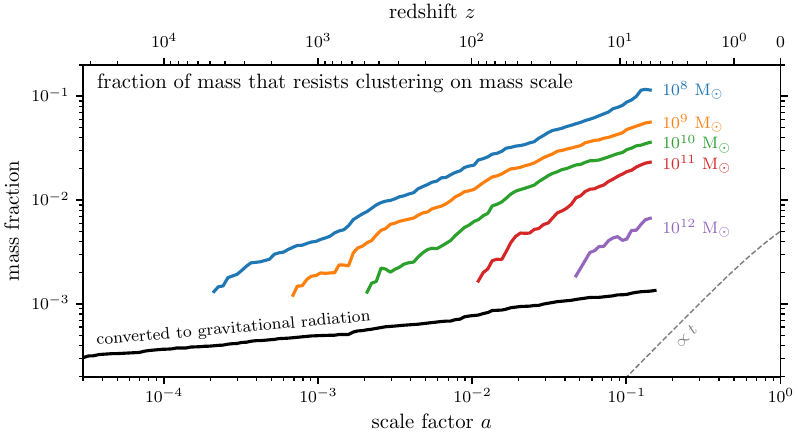}
\caption{
    Fraction of PBH mass that resists clustering due to collisional and relativistic backreactions.
    The colored curves show the fractions that stream across the indicated mass scales due to collisional heating.
    The black curve shows the fraction that has been converted into gravitational radiation and thus does not cluster on any scale (and modifies the cosmic expansion history).
    For comparison, the dashed curve is proportional to time; that is how the decayed fraction would scale for a decaying dark matter model.
    }
\label{fig:totalmass}
\end{figure}

For a range of mass scales $M$, figure~\ref{fig:totalmass} shows the fraction of the dark matter that would resist clustering on the scale $M$ due to its streaming motion (equation~\ref{streaming}).
We also show in this figure a weaker, but more broadly relevant, backreaction: the PBH mass that is converted into gravitational radiation during relativistic interactions, particularly binary merger events.
Around 0.14 percent of the PBH mass is lost in this way by $z\simeq 5.5$.\footnote{For comparison, the PBH count decreases by about 1 percent by the end of the simulation.}
The effect is thus small, but it would affect structure formation at all scales as well as the cosmic expansion history itself.
It is similar in concept to the effect of a decaying dark matter model, but the time dependence is very different. For a small decayed fraction, the mass loss from dark matter decay scales proportionally with time, and we show an example of this scaling (dashed curve). PBH mass decays much more gradually than decaying dark matter.
We also note that this decay of PBH mass produces a substantial stochastic gravitational-wave background, which we will quantify in section~\ref{sec:GW}.

Finally, although the results of this section have been specific to our choice of $\mathcal{O}(10~\Msol)$ PBHs, we can estimate how they should scale with different PBH masses.
If PBHs have mass $m$, then their separations scale as $r\propto m^{1/3}$, so binary orbital velocities scale as $v\propto\sqrt{Gm/r}\propto m^{1/3}$. Collisional heating would produce comparable streaming velocities. Comparing to equations (\ref{streaming}) and~(\ref{escape}), this means that mass scales $M$ at which structure is suppressed by collisional heating scale proportionally to the PBH mass, i.e., $M\propto m$. This is a natural scaling to expect and simply reflects the scale invariance of Newtonian gravity.
Meanwhile, the conversion fraction of PBH mass into gravitational radiation depends on the rate of binary formation and collisional hardening, which are set by Newtonian gravity and hence should be independent of the PBH mass.
However, the radiation conversion may retain some mass dependence through the relativistic orbital decay rate, which scales strongly with PBH mass.
Although we will see in section~\ref{sec:mergers} that relativistic orbital decay does not dominantly drive binary coalescence, it is expected to play some role, and so the conversion rate of PBH mass into gravitational radiation is likely to have some weak but nonzero sensitivity to the PBH mass scale.

\subsection{Collisional halo evolution}

\begin{figure}
  \includegraphics[width=\textwidth]{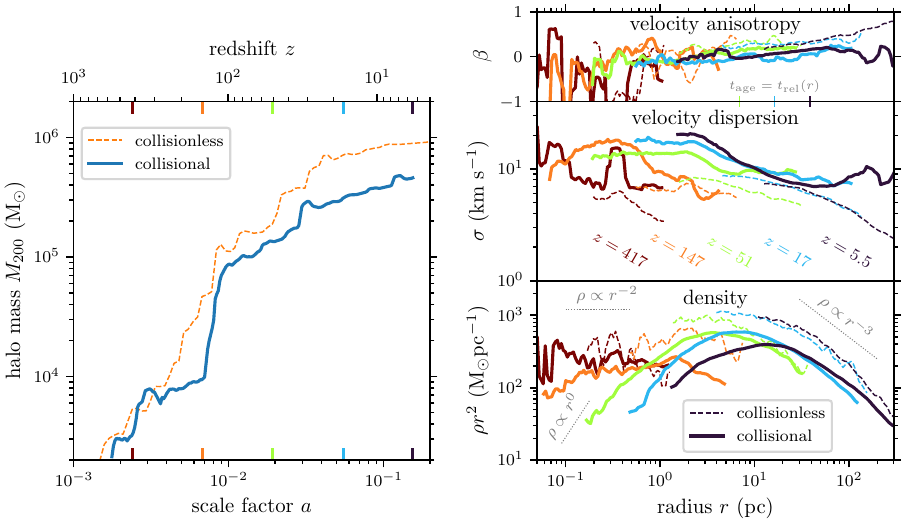}
\caption{
Comparison of halo evolution between the collisional PBH simulation (solid curves) and the collisionless version (dashed curves). The left-hand panel shows the mass accretion history of the largest halo in the simulation volume, while the right-hand panels show the internal structure for a range of times (different colors, corresponding to the tick marks in the left-hand panel). From bottom to top, we show the density profile (scaled by $r^2$ to reduce the dynamic range), the velocity dispersion $\sigma$, and the velocity anisotropy parameter $\beta$. Each is averaged over snapshots covering a factor of about 1.2 in the scale factor $a$ and plotted up to the radius $R_{200}$ enclosing average density 200 times the cosmological mean dark matter density. For the collisional simulation, we plot down to a minimum radius that encloses 100 PBHs across all ($\sim 100$-$1000$) snapshots included in the time average, which typically encloses an average of a few PBHs. For the collisionless simulation, we plot down to 3 times the softening length, since this is about the distance below which forces become artificially non-Newtonian.
}
\label{fig:halo_comparison}
\end{figure}

We turn our attention next to the internal structure of the largest halo in the PBH simulation and how it is affected by collisional dynamics.
Figure~\ref{fig:halo_comparison} shows a comparison between that halo in the collisional PBH simulation and its counterpart in the collisionless simulation.
We consider the growth history and the evolution of density and velocity profiles.
For the velocity profiles, we evaluate the radial velocity dispersion as $\sigma_\mathrm{r}^2=\langle v_\mathrm{r}^2\rangle-\langle v_\mathrm{r}\rangle^2$ and the tangential velocity dispersion as $\sigma_\mathrm{t}^2=\langle \vec v^2\rangle-\langle v_\mathrm{r}^2\rangle$, where $\vec v$ is a particle's velocity, $v_\mathrm{r}$ is the radial component thereof, and the angle brackets average over all particles in a radius bin. We show in figure~\ref{fig:halo_comparison} the total velocity dispersion, $\sigma^2=\sigma_\mathrm{r}^2+\sigma_\mathrm{t}^2$, and the velocity anisotropy parameter, $\beta\equiv1-\sigma_\mathrm{t}^2/(2\sigma_\mathrm{r}^2)$.
Note that $\beta=1$ means all motion is radial, $\beta=0$ means the velocity dispersion is isotropic, and $\beta<0$ means that velocities are preferentially tangential.
These velocity profiles, as well as the density profiles, are evaluated about the minimum of the softened gravitational potential; we employ softening for this purpose even in the collisional simulation because the true potential minimum typically corresponds to a tight binary system that is not necessarily at the center of the halo.
We find this softened potential minimum using the \textsc{subfind} algorithm \cite{Springel:2000qu} as implemented in \textsc{Gadget-4}, and we also use the merger tree algorithm in \textsc{Gadget-4} to trace the halo through time.

At $z\simeq 400$, the halo is comparable in mass and density profile between the collisional and collisionless simulations. The density profile is close to that of an isothermal sphere, $\rho\propto r^{-2}$.
However, over time, the growth of the halo in the collisional PBH simulation is delayed compared to its collisionless counterpart, as we discussed earlier. This delay leads to the halo in the collisional simulation having slightly lower density at most radii.
Throughout its history, the collisional halo has a velocity dispersion almost uniformly higher than that of the collisionless halo. However, this comparison does not have a straightforward interpretation, as some of the velocity dispersion in the collisional halo comes from the orbital motion of binary pairs.
Similarly, the collisional halo has a much more isotropic velocity distribution at large radii than does the collisionless halo, and this may be due to isotropically oriented binary pairs.
Although collisions are expected to isotropize the velocities, this effect is only expected to be significant at small radii, where collisions are frequent (as we quantify below).

\begin{figure}
  \centering
  \includegraphics[width=\textwidth]{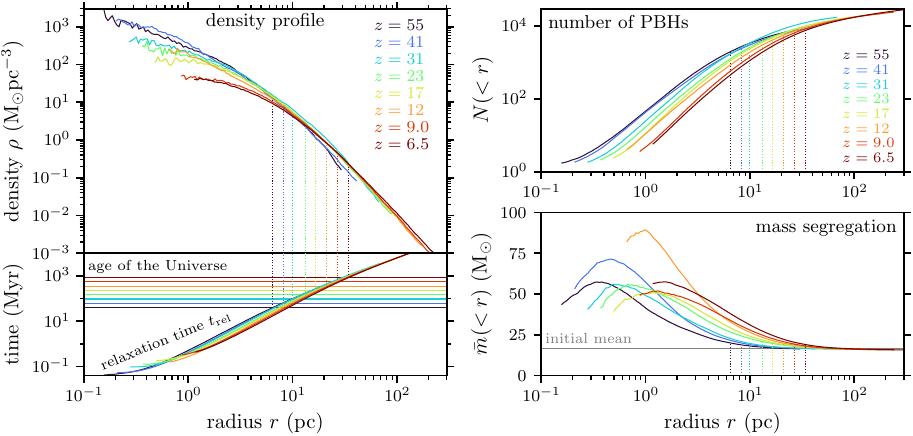}
\caption{Collisional evolution of the largest halo in the simulation volume. The upper left panel shows the density profile for a range of times, similarly to figure~\ref{fig:halo_comparison}. Below it, for the same range of times, we compare the collisional relaxation time to the age of the Universe. Relaxation-induced formation of a finite-density core becomes significant when the age of the Universe is around an order of magnitude longer than the relaxation time scale.
The upper right panel shows the number of PBHs within each radius in the halo, while the lower right panel shows the average mass of PBHs below each radius. Collisional mass segregation becomes significant already when the age of the Universe equals the relaxation time scale (vertical lines).
}
\label{fig:evolution}
\end{figure}

Most strikingly, the collisional halo forms a finite-density core at small radii, which grows in size and decreases in density over time.
The velocity dispersion in the core is a few times higher than elsewhere in the halo; the growth of the density core in figure~\ref{fig:halo_comparison} aligns with growth of the high-velocity-dispersion part of the halo in the same figure.
Core formation is an expected consequence of the exchange of energy in two-body collisions between PBHs, and we explore this effect further in figure~\ref{fig:evolution}.
We define the time scale for two-body collisional relaxation as 
\begin{equation}
    t_\mathrm{rel}(r)\equiv \frac{N(r)}{8\ln N(r)}\sqrt{\frac{r^3}{G M(r)}}
\end{equation}
(e.g.~\cite{2008gady.book.....B}), which we take to be a function of radius $r$ within the halo. Here $M(r)$ and $N(r)$ are the mass and PBH count, respectively, enclosed within the radius $r$. The interpretation is that over the time $t_\mathrm{rel}(r)$, two-body encounters are expected to effect $\mathcal{O}(1)$ changes in the velocities of PBHs below the radius $r$.
In figure~\ref{fig:evolution} (lower left panel), for a range of times, we compare $t_\mathrm{rel}(r)$ to the age of the Universe.

The evolution of the halo density profile in the collisional simulation, shown in figure~\ref{fig:evolution}, can be interpreted as follows. The vertical lines indicate for each time the radius $r$ at which the age of the Universe $t_\mathrm{age}$ equals the relaxation time $t_\mathrm{rel}(r)$. Apparently, collisional relaxation requires around an order of magnitude longer than the relaxation time, i.e.
$t_\mathrm{age}\gtrsim 10 t_\mathrm{rel}(r)$,
to significantly reduce the halo density at the radius $r$.
We also mark the same $t_\mathrm{age}=t_\mathrm{rel}(r)$ radius with tick marks (only for $z\leq 55$) in figure~\ref{fig:halo_comparison}, which shows the velocity distribution. This confirms that major differences in the velocity structure of the collisional halo, compared to the collisionless version, appear outside the regime where relaxation is important, suggesting that they are indeed due to binaries.

Notably, the core only becomes more diffuse over time; it does not begin to collapse.
Core collapse has been suggested to be an important phase in the evolution of PBH halos \cite{Vaskonen:2019jpv,Stasenko:2023zmf,Stasenko:2024pzd}, so it is interesting that its onset does not take place within the duration of our simulation.
During core collapse, heat exchange between a warm core and the colder outskirts of the system causes the core to lose energy, making it contract and become hotter due to its negative heat capacity \cite{Lynden-Bell:1968eqn}. The process accelerates itself as the now hotter core transfers heat outward more rapidly.
Core collapse thus arises when the velocity dispersion decreases with distance \cite{2008gady.book.....B}.
However, as figure~\ref{fig:halo_comparison} shows, the velocity dispersion in our PBH halo does not decrease at large radii, instead remaining close to uniform, likely due to the contribution of binary systems.
This property may explain why core collapse is avoided.

Mass segregation is another important consequence of collisional dynamics.
Two-body collisions tend to transfer energy from heavier to lighter PBHs, making the heavier PBHs sink to the center of the system while the lighter PBHs are elevated.
On the right-hand side of figure~\ref{fig:evolution}, we show how the PBH count $N(r)$ and the average PBH mass $\bar m(r)$ vary as a function of the enclosing radius $r$.
Both quantities begin to evolve significantly, with $N$ decreasing and $\bar m$ increasing, when
$t_\mathrm{age}\gtrsim t_\mathrm{rel}(r)$.
That is, mass segregation becomes important over a time scale of order $t_\mathrm{rel}$, even though significant changes in the overall density take an order of magnitude longer.
This outcome confirms that $t_\mathrm{rel}$ is an appropriate measure of when collisional energy exchange becomes important.

\begin{figure}
\centering
\includegraphics[width=\textwidth]{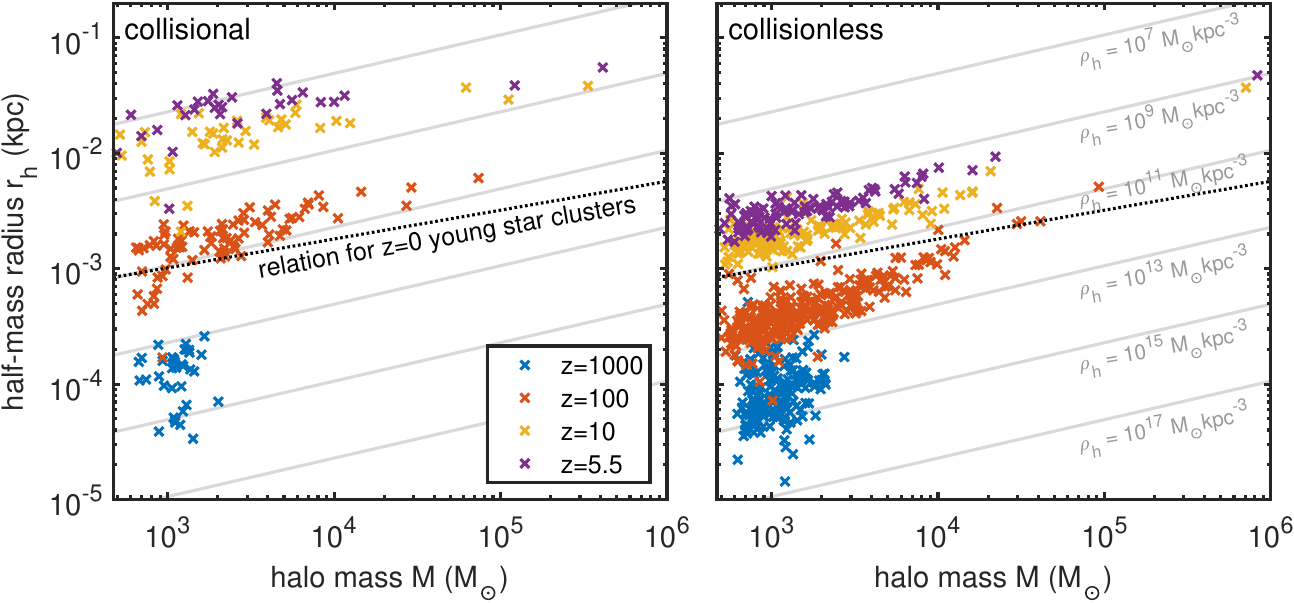}
\caption{Mass-size relation for the PBH halos at four different redshifts from $z=1000$ to $z=5.5$. Lines of constant half-mass density are shown in grey, and the $z=0$ relation for young star clusters is displayed as a dark dotted line.
The left-hand panel corresponds to the collisional PBH simulation, while the right-hand panel shows the collisionless version for comparison.
The halos in the collisionless simulation remain considerably more compact than the collisional PBH halos.
}
\label{fig:masssize}
\end{figure}

Figure~\ref{fig:masssize} shows a broader picture of the collisional evolution of the PBH halos in our simulation.
We consider all gravitationally self-bound clumps of at least 20 PBHs, identified using the \textsc{subfind} algorithm (as implemented in \textsc{Gadget-4}), and we show their distribution in total bound mass $M$ and half-mass radius $r_\mathrm{h}$ (which encloses mass $M/2$).\footnote{\label{foot:subfind}For the collisional PBH simulation, we properly take the gravitational softening length to be 0 for this analysis. Since self-potentials diverge without softening, we modified \textsc{Gadget-4} to ensure that self-potentials are never evaluated. To avoid inappropriately centering on a hard binary, the half-mass radius is evaluated about the center of mass instead of the potential minimum.}
For comparison, we also show the halo distribution in the collisionless simulation.
At $z=1000$, the PBH halos in the collisional and collisionless simulations are comparable in mass and size.
However, at later times, the collisional halos are much more diffuse than their collisionless counterparts.
Indeed, the half-mass density
\begin{equation}
    \rho_\mathrm{h}\equiv\frac{M/2}{(4\pi/3)r_\mathrm{h}^3}    
\end{equation}
of the collisional halos drops over time in approximate proportion with the density of the Universe, $\rho_\mathrm{h}\propto (1+z)^3$, such that these halos remain uniformly about $10^3$ times denser than the Universe at large.
In contrast, the decrease in half-mass density of the collisionless halos is much more gradual. The latter outcome reflects that halos of collisionless dark matter are strongly sensitive to their growth history, with early-forming halos being internally denser (e.g.~\cite{Dalal:2010hy,Ludlow:2013bd,Delos:2019mxl,Delos:2022yhn}). Apparently, the collisional evolution of the PBH halos tends to erase the influence of this history.

\subsection{Comparing PBH clusters to star clusters}

From the dynamical point of view, halos of collisional PBHs closely resemble clusters of stars in the point-mass particle limit \cite{Aarseth2003}. For PBHs, the complexities of the single and binary stellar evolution can be conveniently omitted, and the Newtonian and post-Newtonian gravitational dynamics completely determine the internal evolution of the PBH halos. However, there are a number of important differences which we elaborate on in this section. We focus on the mass-size relation of the halos, pictured in figure~\ref{fig:masssize}, which determines their relaxation and mass segregation times.
We also examine the binary PBH population properties in comparison to simple binary star population models.

While star clusters form from collapsing and fragmenting clouds of low-temperature gas, PBH halos form cosmologically from collapsing overdense regions. Most importantly, PBH halos continue evolving due to accretion of PBHs onto the halos at later times.
While star clusters can grow by a factor of $\sim10$ from their small birth radii during their early evolution, figure~\ref{fig:masssize} shows that the PBH halo mass-size relation strongly evolves as a function of redshift.
Early PBH halos ($z\sim1000$) have masses of order $10^3~\Msol$ and half-mass sizes of order $0.1$ kpc, making them approximately an order of magnitude more compact than typical young star clusters \cite{Brown2021} of the same mass. 

Dark matter halos grow in mass and spatial size through ongoing accretion from
their environments. Additionally, compared to halos of collisionless matter, the size growth of collisional PBH halos is more rapid due to two-body and binary processes. The PBH halos reach the mass-size relation for young star clusters somewhat before $z=100$, which reflects that only at very early times does the mean density of the Universe compare to that of star-forming environments.
By $z=5.5$, the PBH clusters are larger in size than young star clusters of comparable mass by a factor of around $10$ to $50$.

The small spatial sizes at high $z$ imply very short mass segregation timescales for the halos. The mass segregation timescale is also shortened due to the large average PBH mass of $16.5~\Msol$ compared to $\sim0.6~\Msol$ for an average star from a typical initial mass function. Thus, the PBH halos are mass segregated already at early times, the most massive PBHs residing at the central regions of the halos. Towards later times, PBH halos grow by hierarchical merging and accretion, increasing their mass and spatial size. By $z=10$, the halos are already over an order of magnitude more extended than typical young star clusters with the trend continuing towards smaller redshifts.

\section{Binary population and gravitational waves}\label{sec:binaries}

Since the first post-LIGO studies of PBH dark matter \cite{Bird:2016dcv,Clesse:2016vqa,Blinnikov:2016bxu}, calculations were performed to include the direct formation of binaries in the very early universe, around the time of matter-radiation equality \cite{Bringmann:2018mxj,Raidal:2018bbj,Ballesteros:2018swv,Liu:2018ess,Raidal:2017mfl,Chen:2018czv,Ali-Haimoud:2017rtz,Sasaki:2016jop}. Mergers of these \emph{early} binaries are predicted to dominate the overall merger rate.
The current consensus is that the observed merger rate rules out that stellar-mass PBHs make up the entirety of the dark matter, as this scenario would lead to too many detectable binary black hole mergers.
Current constraints imply that, at most, stellar mass PBHs could compose $\mathcal{O}(0.1\%)$ of dark matter (where constraints depend somewhat on the PBH mass function), i.e., $f_{\rm PBH} \lesssim 0.001$, 
where $f_\mathrm{PBH}$ is the fraction of dark matter composed of PBHs.

However, there is significant uncertainty in applying constraints when $f_{\rm PBH}$ becomes larger, as interactions of early binaries with other nearby PBHs can alter the orbits of early binaries, or prevent them forming in the first place \cite{Raidal:2018bbj}. Several attempts have been made to account for the interactions of binary PBHs with other PBHs. Jedamzik \cite{Jedamzik:2020omx,Jedamzik:2020ypm} studied binaries in early, extremely dense, halos, finding that any binaries in such halos would almost certainly be disrupted, and thus not contribute to the merger rate today, although the fraction of binaries which end up in such systems was uncertain. Young \& Hamers \cite{Young:2020scc} studied binaries in \emph{Milky Way}-type halos, finding that, since such halos are typically much less dense, the orbits of binaries are unlikely to be significantly affected---although this neglects effects such as mass segregation and PBH clustering inside galactic halos.

Motivated by these considerations, the principal focus of this section is to study (i) whether accurate predictions can be made for the initial orbital parameters of early binaries using existing calculations and (ii) whether the initial conditions of a binary provide a good estimate for when it will merge.
We will also estimate the gravitational-wave background produced by PBH mergers.
A fuller analysis of binary properties and mergers will be pursued in a follow-up paper.

For our analysis, we consider two PBHs to compose a binary system if and when the pair have negative gravitational potential energy, and we only include PBH binaries which are not strongly perturbed by third bodies. We quantify the relative strength of the tidal perturbation through the parameter $\gamma_\mathrm{pert}$ at the binary apocenter defined as
\begin{equation}\label{eq:gammapert}
\gamma_\mathrm{pert} =  \frac{m_\mathrm{3}}{r_\mathrm{3}^3} \left\{\frac{m_\mathrm{1} m_\mathrm{2}}{(m_1+m_2) \left[r_\mathrm{a}(1+e) \right]^3}\right\}^{-1}.
\end{equation}  
Here $m_\mathrm{3}$ is the mass of the closest PBH to the binary and $r_\mathrm{3}$ is its separation from the binary center-of-mass. We only consider binaries with $\gamma_\mathrm{pert}<0.3$.

We will further separate the binaries into 3 categories, based upon the initial positions of the PBHs:
\begin{itemize}
    \item Early binaries are classed as those which are initially each other's nearest neighbour at the start of the simulation, and these typically form at very early times. Approximately 12\,000 early binaries are identified.
    \item Late binaries are classed as those where neither is initially the nearest neighbour of the other, and these typically form at significantly later times.
    Approximately 20\,000 late binaries are identified.
    \item Finally, there is a population of binaries which do not fit easily into either category, which we classify as ambiguous binaries. This can be the case where either one PBH was the nearest neighbour to the second, but the second PBH was initially closer to a third, or early binaries which became temporarily unbound, before once more forming a binary system. Approximately 8\,000 ambiguous binaries are identified.
\end{itemize}
These binary counts are across all times; the number of binaries at any one time is much lower, as we will see shortly.
Although we do not attempt to identify the formation channels of the PBH binaries, the general expectation is that early binaries form in the field when they decouple from the Hubble flow \cite{Bringmann:2018mxj,Raidal:2018bbj,Ballesteros:2018swv,Raidal:2017mfl,Chen:2018czv,Ali-Haimoud:2017rtz,Sasaki:2016jop}, while late binaries form inside virialized halos either by many-body PBH interactions \cite{Franciolini:2022ewd,Stasenko:2024pzd} or by capture induced by the gravitational radiation reaction force \cite{Bird:2016dcv,Clesse:2016vqa,Blinnikov:2016bxu}.

\subsection{Number and hardness of PBH binaries}\label{sec:binarycount}

\begin{figure}
\centering
\includegraphics[width=\textwidth]{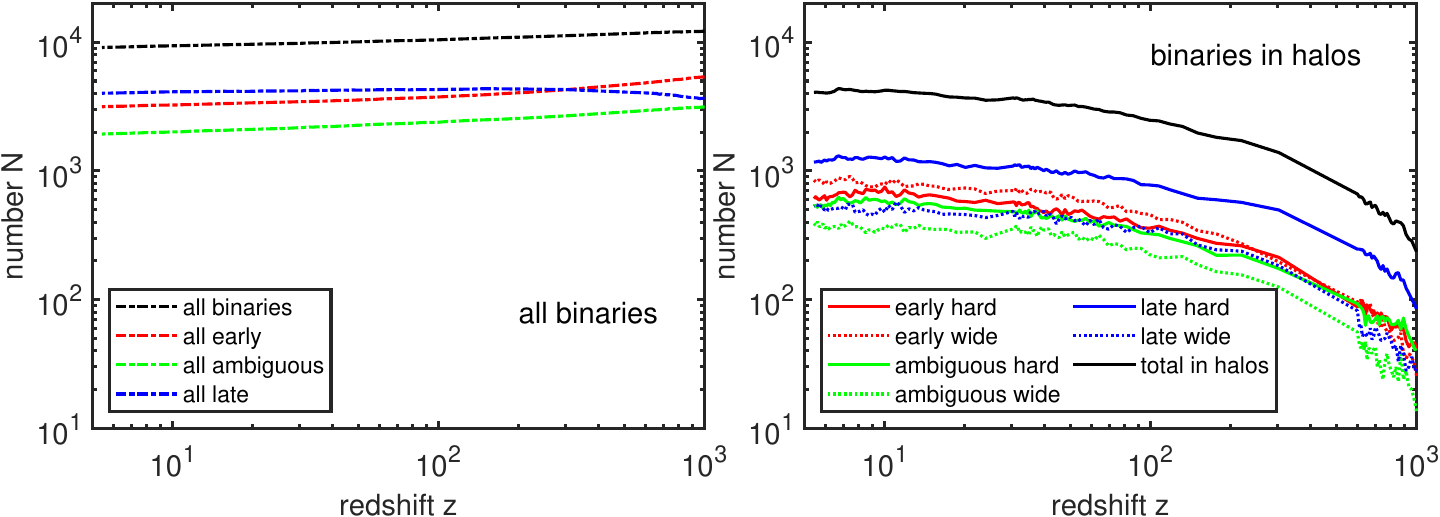}
\caption{
Number of PBH binaries over time, plotted as a function of redshift.
The left-hand panel shows the total number of PBH binaries, as well as the number that are classified as early, late, or ambiguous.
The right-hand panel shows the number of PBHs that reside in bound clusters identified by the \textsc{subfind} algorithm. Here we additionally distinguish hard binaries (which satisfy equation~\ref{wide-hard-definition}) from wide binaries.
}
\label{fig:binaryhardness}
\end{figure}

We begin with a summary of the number of PBH binaries in our simulation and the dynamical status of these systems with respect to their environments.
Figure~\ref{fig:binaryhardness} shows how the number of binaries changes over time.
Approximately 12\,000 binaries are present in the simulation by $z\sim 1000$, implying that about 30 percent of all PBHs are in binaries. By the end of the simulation at $z\simeq 5.5$, the binary count decreases only moderately to about 9\,000, a number that accounts for about 23 percent of the PBHs. This decline is mostly in the early and ambiguous binaries; the number of late binaries remains fairly steady after $z\sim 700$.

We also show how many of these binaries reside in bound clusters of PBHs, as identified by the \textsc{subfind} algorithm.\footnote{We take the softening length in \textsc{subfind} to be zero, as discussed in footnote~\ref{foot:subfind}.}
Over the same duration from $z=1000$ to $z\simeq 5.5$, the fraction of binaries that reside in bound clusters rises from 2--3 percent to about 47 percent.\footnote{Many of the PBH binaries not in clusters at late times were ejected from them, as we found in section~\ref{sec:heating}. Consequently, the fraction of binaries not in clusters at late times is very different from the fraction of binaries that can be taken to have dynamically unperturbed orbits, as considered by refs.~\cite{Vaskonen:2019jpv,DeLuca:2020jug} for the purpose of estimating the PBH merger rate.}
A higher proportion of late binaries reside in clusters, compared to early binaries. This naturally reflects the expectation that late binaries form inside clusters while early binaries form in the field.

Binaries residing in bound clusters can be characterized as \emph{hard} or \emph{wide}.
According the so-called Heggie-Hills law \cite{Heggie1975,Hills1975}, hard (or close) binaries tend to further harden in interactions with other cluster bodies, becoming more tightly bound over time.
Meanwhile, wide binaries on average expand and eventually dissolve. A simple way to determine whether a PBH binary is hard or wide in its environment is to compare its binding energy $B=-E=-T-U$ to the mean kinetic energy $T$ of passing PBHs. Using the virial theorem, $B=T=1/2 \mu v_\mathrm{orb}^2$ in which $\mu=m_\mathrm{1}m_\mathrm{2}/(m_1+m_2)$ is the reduced mass of the binary and $v_\mathrm{orb} = \sqrt{G (m_1+m_2)/r_\mathrm{a}}$ its circular orbital velocity. A mean passing PBH has a kinetic energy of $\bar m \sigma^2/2$ where $\bar m$ is the mean PBH mass and $\sigma$ is the velocity dispersion of the cluster that the binary resides in. The resulting criterion for binary hardness is
\begin{equation}\label{wide-hard-definition}
\left( \frac{\mu}{\bar m}\right)^{1/2} v_\mathrm{orb}>\sigma.
\end{equation}
We obtain the velocity dispersions from the \textsc{subfind} data.

Figure~\ref{fig:binaryhardness} shows how the counts of hard or wide binaries evolve with time in our simulation.
About 20 percent of all binaries are wide, and thus expected to dissolve eventually, at the end of the simulation at $z\simeq 5.5$. The rest are either hard or not residing inside a bound cluster.
Within clusters, early binaries tend to be predominantly wide, while late binaries tend to be predominantly hard.
This outcome likely represents survivorship bias resulting from where these binaries form.
Late binaries tend to form near the dense centers of clusters, where interactions are frequent. There, wide binaries would rapidly dissolve, leaving only hard binaries. Conversely, early binaries tend to form in the field, and those inside clusters would often reside in the less dense outskirts, where interactions are rare. These binaries could often persist for a long time, even as wide binaries.

\begin{figure}
\centering
\includegraphics[width=0.88\textwidth]{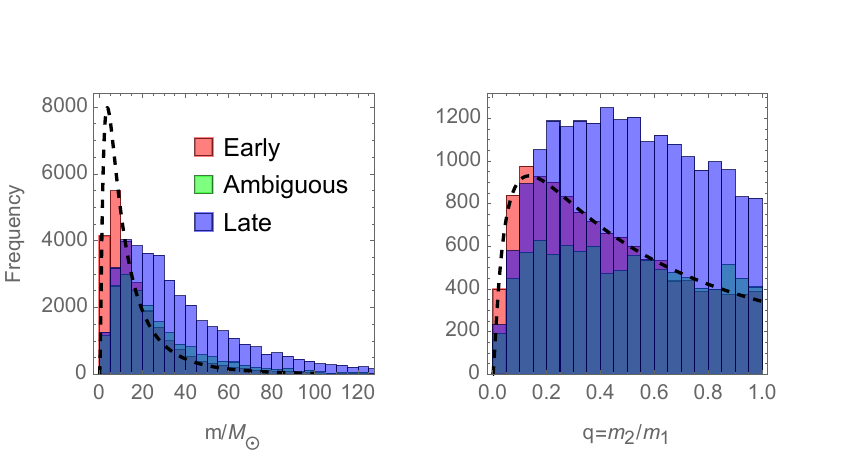}
\caption{
Distribution of binaries in terms of the masses of member PBHs (left-hand panel) and the mass ratio (right-hand panel).
PBHs in late binaries tend to be heavier than those in early or ambiguous binaries, likely reflecting the tendency of late binaries to form in the mass-segregated centers of clusters of PBHs.
For comparison, the dashed line in the left-hand panel shows the underlying mass distribution of the PBHs (equation~\ref{PBH_mass-function}), scaled to the number of early binaries.
In the right-hand panel, the dashed line shows the distribution of mass ratios that would result if early binaries arise from independent pairwise draws from the PBH distribution (equation~\ref{mass-ratios}).
}
\label{fig:binarymasses}
\end{figure}

Figure~\ref{fig:binarymasses} shows the mass distribution of the PBHs that belong to binaries identified in the simulation. Late binaries are systematically heavier than early binaries, likely a consequence of their tendency to form in the dense centers of PBH clusters, where the PBHs are more massive due to mass segregation (as illustrated in, e.g., figure~\ref{fig:evolution}).
For early binaries, the mass distribution more closely resembles the underlying lognormal mass distribution of the PBHs (equation~\ref{PBH_mass-function}), although there is a modest deficit of low-mass early binaries.
This deficit likely arises because low-mass PBH pairs are less likely to form weakly perturbed binaries in accordance with equation~(\ref{eq:gammapert}).

Figure~\ref{fig:binarymasses} also shows the distribution of binary mass ratios $q=m_2/m_1$, where $m_1>m_2$. Early binaries arise from random pairing in the initial conditions, so the distribution of their mass ratios is simply the ratio of two random draws from the PBH mass distribution given by equation~(\ref{PBH_mass-function}). Since the mass distribution is lognormal, the ratio distribution for random pairs is also lognormal and has the form
\begin{align}\label{mass-ratios}
    f(q)
    &=
    \frac{\e^{-(\ln q)^2/(4\sigma_{\ln}^2)}}{\sqrt{\pi}\,\sigma_{\ln} q}
\end{align}
(recall we set the width of the mass function to be $\sigma_{\ln}=1$). Figure~\ref{fig:binarymasses} shows that equation~(\ref{mass-ratios}) closely resembles the mass ratio distribution for early binaries in the simulation. In contrast, the members of late binaries tend to have more equal masses, likely due to the expulsion of all low-mass PBHs from the cores of the PBH clusters.

\subsection{Initial distribution of binary orbits}\label{sec:binaryorbits}

\begin{figure}
\centering
\includegraphics[width=0.85\textwidth]{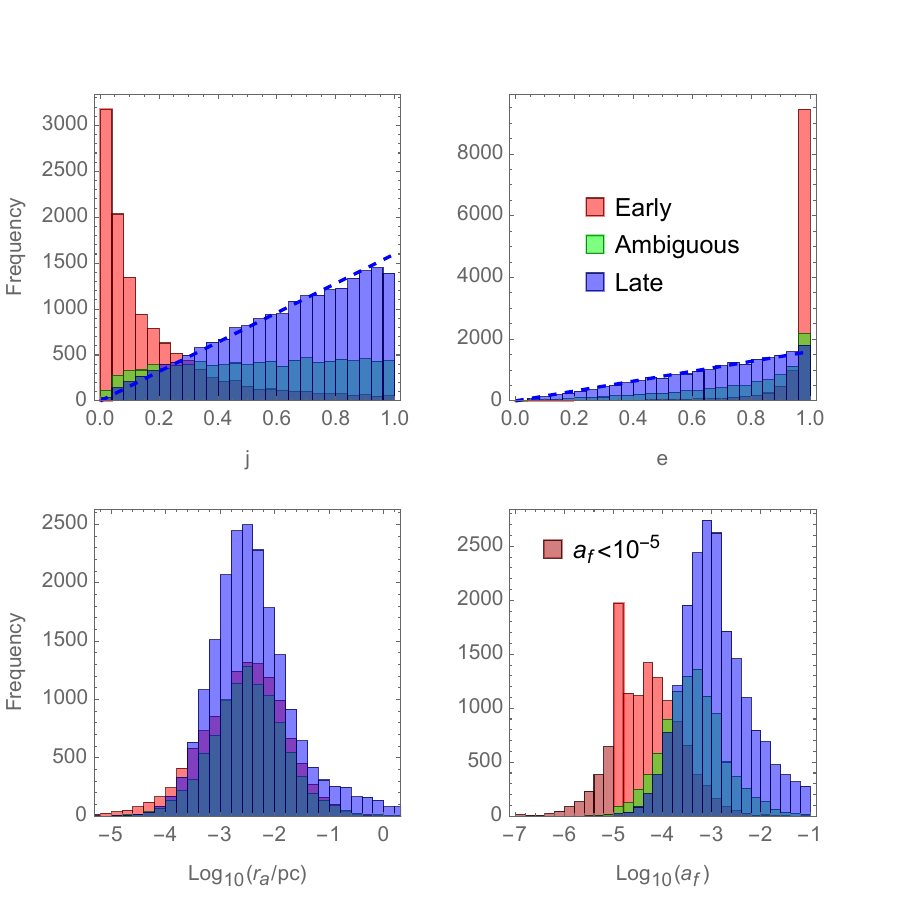}
\caption{Distributions of initial orbital elements of PBH binaries. We show the dimensionless angular momentum $j$ (upper left), eccentricity $e$ (upper right), semi-major axis $r_\mathrm{a}$ (lower left), and scale factor $a_\mathrm{f}$ of binary formation (lower right), all evaluated when the binary is first identified in the simulation. 
For the $j$ and $e$ distributions, the dashed blue lines show the expected values for a thermal distribution.
For the $a_\mathrm{f}$ distribution, the darker red color indicates early binaries identified at $a_\mathrm{f}<10^{-5}$. The spike in binary formation at $a_\mathrm{f}=10^{-5}$ is due to the different criteria used to identify binaries before and after this time -- and we note that the formation time of binaries should be treated as approximate, as it is dependent on the (somewhat) arbitrary criteria used.
}
\label{fig:binaryelements}
\end{figure}

We now turn our attention to the distribution of binary orbits, shown in figure~\ref{fig:binaryelements}.
We evaluate the orbital elements at the time that each binary was first identified in the simulation, although binary orbits evolve over time, as we will see later.
We also show the distribution of these binary identification times. Before $a=10^{-5}$, we identify a binary at the time of its first pericenter passage. Afterward, we identify a binary as soon as the binding energy and $\gamma_\mathrm{pert}$ criteria are satisfied.

Here, we will focus on determining whether the early binaries identified in the simulation match those predicted by the analytic model of Raidal, Spethmann, Vaskonen and Veerm\"ae \cite{Raidal:2018bbj}, hereafter referred to as RSVV. The model provides a framework for an analytic prediction of the number of binaries based on the abundance and mass function of PBHs, including calculations for the expected semi-major axes and eccentricities. They find that the predictions are in excellent agreement with some simple simulations of early binary formation, but that there are significant uncertainties when the PBH fraction is large. Such calculations are essential in order to place constraints on the PBH fraction from the gravitational-wave signals of merging black holes observed by LVK. 

Within the framework of RSVV, a binary is considered to form if two PBHs are initially close enough to each other that their self-gravitation overcomes the Hubble flow, and if there are no other nearby PBHs in a sphere of radius $y$ (known as the exclusion radius). To parameterise the uncertainty in whether a binary forms, this radius is treated as an unknown for most of the calculation; more specifically, the expected PBH count $N(y)=\bar n V(y)$ inside a sphere of radius $y$ is considered, where $\bar n$ is the mean PBH number density and $V(y)=4\pi y^3/3$ is the volume.

\begin{figure}
  \begin{centering}
  \includegraphics[width=0.6\textwidth]{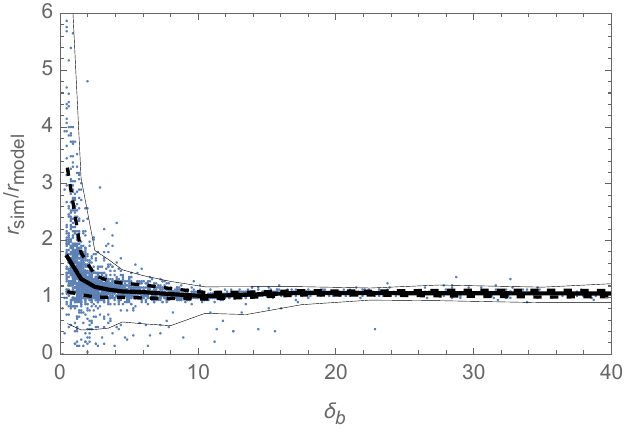}
\caption{
Comparing the initial semi-major axes $r_\mathrm{sim}$ of early binaries that form in the simulation to the values $r_\mathrm{model}$ that are predicted by equation~\eqref{eqn:raModel}. We show this comparison as a function of $\delta_b$, the initial overdensity associated with the binary. One out of every hundred binaries is plotted with a blue point, while the thick black line represents the median value in each bin. The dotted and thin black lines enclose the middle $68\%$ and $95\%$ of the data in each bin, respectively. For large $\delta_b$, the analytic model of RSVV provides a very good prediction for the semi-major axis, but there is a large deviation from this model for small $\delta_b$}.
\label{fig:raModelComparison}
\end{centering}
\end{figure}

For the binaries which might be expected to merge within the age of the Universe, the initial pair with comoving separation $x_0$ represents a matter overdensity described by the parameter
\begin{equation}
    \delta_b = \frac{m_1+m_2}{2 \rho_\mathrm{M} V(x_0)},
\end{equation}
where $\rho_\mathrm{M}$ is the mean total matter density. The pair decouples from the Hubble flow when the matter density exceeds the radiation density within the region, which sets the semi-major axis $r_\mathrm{a}$ for the newly formed binary,
\begin{equation}
    r_\mathrm{a} \approx 0.1 \frac{a_\mathrm{eq}x_0}{\delta_b},
    \label{eqn:raModel}
\end{equation}
where $a_\mathrm{eq}$ is the scale factor at matter-radiation equality. For early binaries identified in the simulation, figure \ref{fig:raModelComparison} compares the semi-major axis calculated with this equation, using the initial conditions of the simulation, with the actual semi-major axis found in the simulation. For initial pairs with large $\delta_b$, equation \eqref{eqn:raModel} can therefore be seen as a good indicator for the semi-major axis. However, for pairs with small $\delta_b$, it is not a good estimator. This is because, for small $\delta_b$, the binary takes a longer time to decouple from the Hubble flow and thus has more time to be influenced by other nearby PBHs. For binaries expected to merge today, we find $\delta_b\gtrsim 5$, which suggests that equation \eqref{eqn:raModel} is valid for calculating $r_\mathrm{a}$.

Figure \ref{fig:binaryelements} shows the distribution of semi-major axes for the early, late, and ambiguous binaries identified in the simulation.
As noted above, these are evaluated when the binary was first identified in the simulation.
It is notable that the distributions of semi-major axis for each of the classes of binary are very similar.
Some correspondence between the different populations is expected, because binary separations should scale proportionally with $\aeq n^{-1/3}$, the characteristic PBH separation at the time of matter-radiation equality.
However, it is surprising that there is essentially no offset between the semi-major axis distributions of the early and late binaries, especially given their very different formation times (also shown in figure~\ref{fig:binaryelements}).

We will now turn our attention to the shape of the binary orbit, described by its eccentricity $e$, although it will be more convenient for us to use the dimensionless angular momentum $j$, related to the eccentricity as
\begin{equation}
    j = \sqrt{1-e^2}.
\end{equation}
In the absence of any other forces acting on the binary pair, the PBHs, initially moving directly away from each other in the centre of mass frame, would collide head on once decoupled from the Hubble flow. However, because one of the PBHs is likely to be closer to other PBHs (or because of density perturbations, although that effect is calculated to be sub-dominant for large PBH fractions), this will provide a torque to the system, imparting some angular momentum and preventing a head-on collision.
In the framework of RSVV, the characteristic angular momentum $j_0$ is calculated to be given by
\begin{equation}
    j_0 \approx 0.4 \frac{f_\mathrm{PBH}}{\delta_b}.
\end{equation}

\begin{figure}
  \begin{centering}
  \includegraphics[width=0.7\textwidth]{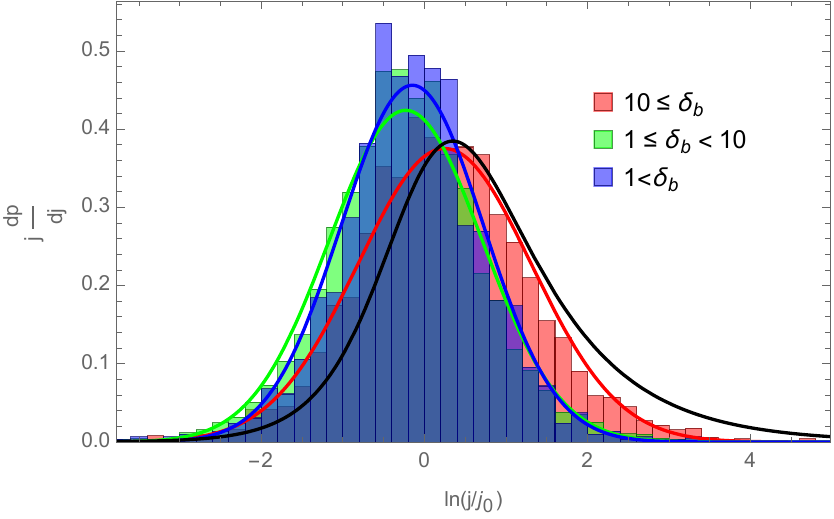}
\caption{Distribution of the dimensionless angular momentum $j$ for early binaries in different ranges of the initial overdensity $\delta_b$. The colored lines represent a fitted lognormal distribution for each case. The black line is the predicted distribution for the limit $N(y)\rightarrow 0$, which we can expect to be valid for $\delta_b \gg 1$.}
\label{fig:jDistributionBifrost}
\end{centering}
\end{figure}

The specific angular momentum of each binary will differ from $j_0$, dependent upon the location and number of nearby PBHs. RSVV therefore calculates a distribution for $j_0$. Here, we will consider the limit of small $N(y)$, expected for binaries with large initial overdensity $\delta_b$, which are initially very close. In this case, the predicted distribution of $j$ is given by a power-law distribution
\begin{equation}
    j \frac{\mathrm d P}{\mathrm d j} = \frac{j^2/j_0^2}{(1+j^2/j_0^2)^{3/2}}.
    \label{eqn:jDist}
\end{equation}
We compare this distribution to the values of $j$ taken from early binaries identified in the simulation, separated by the values of $\delta_b$, in figure~\ref{fig:jDistributionBifrost}. For large values, $\delta_b>10$, we find that equation \eqref{eqn:jDist} is indeed a good fit to the data, although it slightly overpredicts the large-$j$ tail of the distribution. We note that, for smaller $\delta_b$, the calculation of RSVV (evaluated beyond the small-$N(y)$ limit of equation~\ref{eqn:jDist}) is a poor fit to the data, with the large $N(y)$ case significantly underpredicting the tail of the distribution. 


By comparison, figure \ref{fig:binaryelements} shows the distributions of $j$ for the different binary classes.
The differences between the distributions are clearly identifiable in the figure. The early binaries typically have a small angular momentum $j$, while the late binaries follow a thermal distribution, $\mathrm d P / \mathrm d j = 2j$, an expected consequence of interactions with other PBHs \cite{Heggie1975}. The ambiguous binaries, not surprisingly, fall between the two, with an approximately uniform distribution.

This analysis has demonstrated that the analytic calculations of RSVV are valid for calculating the semi-major axis and angular momentum of early binaries with large $\delta_b$. Since binaries expected to merge approximately 13.8 billion years later (i.e. today) typically have a large $\delta_b$, we conclude that the analytic model is suitable for determining the initial parameters of such binaries.

\subsection{Comparing PBH binaries to stellar binaries}

We briefly comment on how the PBH binaries that arise in our simulation compare to simple stellar binary population models.
The distribution of semi-major axes, shown in figure~\ref{fig:binaryelements}, has a roughly log-normal shape with a maximum close to $r_\mathrm{a}\sim10^{-3}$ pc, which is comparable to that of typical dynamically formed stellar binary systems. 
For the early PBH binaries, however, there is a long tail of very tight binaries with much smaller separations, approaching those typical of primordial binary stars (which originate from a common cloud core phase).
Late PBH binaries have a thermal eccentricity distribution with $f(e) \propto e$, which is also typical of stellar binary populations.
However, early PBH binaries have extremely high eccentricities, as discussed in the previous section.

We can also compare the binary mass ratios shown in figure~\ref{fig:binarymasses}.
For ambiguous and late PBH binaries, the distribution $f(q)$ of mass ratios $q$ is surprisingly close to the mass ratio distribution for massive stellar binaries, $f(q) \simeq \text{const}$ for $q>0.1$.
As we noted earlier, these PBHs are expected to pair mostly in the dense, mass-segregated centers of PBH clusters.
The population of early PBH binaries exhibits a larger proportion of more extreme mass ratios, peaking around $q\sim 0.1$ due to random PBH pairing at early times, as discussed in section~\ref{sec:binarycount}.

\subsection{Timings of binary mergers}\label{sec:mergers}

After pairing, the PBHs of a binary will lose energy over time due to gravitational wave emission and spiral inwards, eventually coalescing into a single black hole. If the binary is hard, its coalescence could be accelerated by dynamical interactions, while if it is wide, interactions might prevent it from coalescing altogether. We find approximately 800 binary coalescence events in our simulation, of which only two are second-generation mergers.
Here, we test how well the initial orbit of a binary (discussed in section~\ref{sec:binaryorbits}) predicts whether and when it eventually coalesces.

The coalescence time of a binary system due to gravitational radiation depends on the masses of the PBHs, the semi-major axis, and the angular momentum. For highly eccentric (low $j = \sqrt{1-e^2}$) systems, the coalescence time is
\begin{equation}
    \tau = \frac{3}{85}\frac{c^5}{G^3}\frac{r_\mathrm{a}^4 j^7}{m_1 m_2 (m_1+m_2)}.
\label{eqn:tau}
\end{equation}
Compared to the general expression valid for arbitrary $j$, equation~(\ref{eqn:tau}) is expected to be accurate to $\mathcal{O}(10\%)$ for typical early binary PBHs \cite{Young:2020scc} and may overestimate the coalescence time by a maximum factor of 1.85 for non-eccentric orbits  \cite{Peters1963,Peters1964}. This possible error is not a major concern; we will see that predictions are typically off by orders of magnitude.

\begin{figure}
\begin{centering}
  \includegraphics[width=0.8\textwidth]{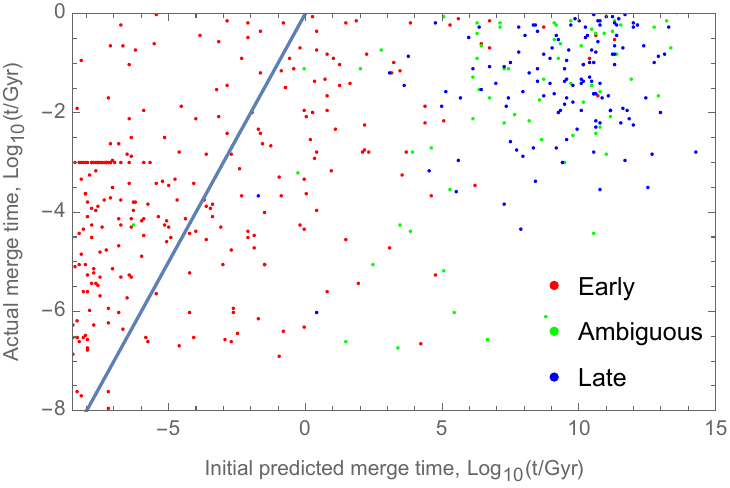}
\caption{Mergers recorded in the simulation plotted against their predicted merge times from initial orbit of each binary. There is remarkably little correlation; an accurate prediction would lie on the diagonal line. The cluster of mergers at 1~Myr arises artificially because the coalescence criterion in the simulation is relaxed at that time.}
\label{fig:recorded_mergers}
\end{centering}
\end{figure}

For binaries observed to have coalesced during the time of the simulation, we can then make a direct comparison between the coalescence time predicted from the initial conditions of the binaries, to the coalescence times observed, with the results shown in figure \ref{fig:recorded_mergers}. We can see that, while a significant number of mergers do occur, there is little correlation between the predicted and observed coalescence times.
The cluster of merger events at 1~Myr arises artificially because the criterion for PBH mergers is relaxed then to speed up the collisional simulation; mergers after 1~Myr are artificially shifted to slightly earlier times.

\begin{figure}
\begin{centering}
  \includegraphics[width=0.8\textwidth]{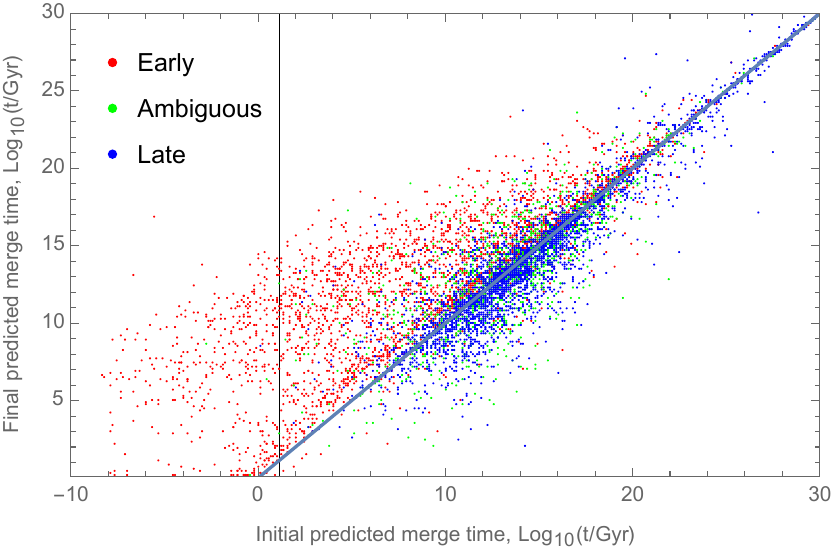}
\caption{Merge times predicted from each binary's orbit at the end of the simulation plotted against those predicted from the binary's initial orbit.
Binaries for which the two predictions agree lie on the diagonal line.
The vertical line represents the current age of the Universe.}
\label{fig:merger_rates}
\end{centering}
\end{figure}

In figure \ref{fig:merger_rates}, we also compare the predicted coalescence times of binaries from their initial conditions with their predicted coalescence times from their orbital parameters at the conclusion of the simulation -- although we note that, when looking at binaries expected to merge e.g. today, this therefore neglects changes to the binaries occurring in the $\sim 12.8$~Gyr after the end of the simulation. For early binaries, there is a weak correlation, and binaries are typically predicted to coalesce at a later time by orders of magnitude compared to the initial prediction. However, for the late and ambiguous binaries, there is a relatively close match between the initial and final predictions -- likely due to their later formation, and thus smaller time available for interactions with other PBHs.

It is interesting to note that equation~(\ref{eqn:tau}) predicts that the vast majority of the late binaries are not expected to coalesce within many orders of magnitude longer than the age of the universe. The initial orbits of the binaries (as indicated in figures \ref{fig:recorded_mergers} and~\ref{fig:merger_rates}) and their orbits at the end of the simulation (as indicated in figure~\ref{fig:merger_rates}) both result in this prediction.
But the prediction is wrong, and figure~\ref{fig:recorded_mergers} shows that numerous late binaries did merge during the simulation.
This may be evidence that mergers of late binaries are dominantly driven by interactions with other PBHs, not primarily through gradual collisional hardening of the semi-major axis, but instead because each collisional interaction results in a new random eccentricity $e$ (drawn from the thermal distribution with $f(e)\propto e$).
Due to the extremely strong sensitivity of the coalescence time in equation~(\ref{eqn:tau}) to the binary eccentricity, coalescence can occur rapidly when a sufficiently high $e$ is drawn.
This stochastic mechanism was noted by ref.~\cite{Rodriguez:2018pss} to be an important channel for mergers of binary black holes in star clusters, and it was also considered by ref.~\cite{Michaely:2019aet} as a possible driver of black-hole mergers in systems of lower density.
It would naturally predict a very poor correlation between a binary's orbit and the timing of its future coalescence.

We conclude therefore that, for large PBH fractions (in this case, $f_\mathrm{PBH}=1$), while we can make a good prediction for the initial abundance of binaries and their orbital parameters, these provide an extremely poor estimate for the times at which the binaries coalesce, and therefore also the merger rate. This outcome agrees with the results of RSVV, who found that their analytical framework is not accurate when $f_\mathrm{PBH}\gtrsim 0.1$.
The reason for the large disagreement between the initial predicted and observed coalescence times in the simulation is the interaction of binaries with other nearby PBHs, which strongly perturb the orbit of the binary. Further study in order to calculate accurate constraints on the PBH abundance arising from the observed LVK merger rate is therefore recommended. A more detailed study of the formation and evolution of binary systems, and the merger rate (and resulting gravitational wave signals) in the simulation, will follow in a second paper.

\subsection{Stochastic gravitational-wave background}
\label{sec:GW}

\begin{figure}
\centering
  \includegraphics[width=\textwidth]{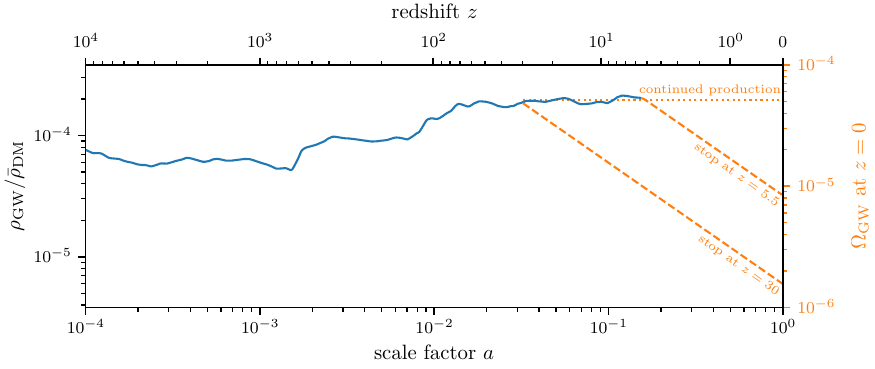}
\caption{
    Energy density $\rho_\mathrm{GW}$ in gravitational radiation compared to the mass density $\bar\rho_\mathrm{DM}$ of the PBHs.
    At late times, the ratio is steady at about $\rho_\mathrm{GW}/\bar\rho_\mathrm{DM}\simeq 2\times 10^{-4}$.
    Extrapolating this behavior up to $z=0$ (dotted line) leads to an energy density parameter of $\Omega_\mathrm{GW}\simeq 5\times 10^{-5}$ today.
    More conservative extrapolations (dashed lines) lead to $\Omega_\mathrm{GW}\sim 10^{-6}$ or $10^{-5}$ by neglecting radiation produced after $z=30$ or $z=5.5$, respectively.
    }
\label{fig:GW}
\end{figure}

Although we postpone a detailed study of binary mergers to a second paper, we can estimate the abundance of gravitational radiation based on the results of section~\ref{sec:backreaction}. Recall that approximately 0.1 percent of the PBH mass was converted into gravitational radiation by the end of the simulation at $z\simeq 5.5$.
We can consider the fraction $f_\mathrm{conv}(a)\ll 1$ of the initial total PBH mass that has been converted into radiation by the scale factor $a$ (black curve in figure~\ref{fig:totalmass}).
The energy density of the radiation is then
\begin{align}
    \rho_\mathrm{GW}(a)
    = 
    \bar\rho_\mathrm{DM}(a)\int_{0}^a\diff a_1 \frac{\diff f_\mathrm{conv}(a_1)}{\diff a_1}\frac{a_1}{a},
\end{align}
where $\bar\rho_\mathrm{DM}(a)$ is the mean mass density in PBHs at $a$.
Figure~\ref{fig:GW} shows the evolution of $\rho_\mathrm{GW}$. By the end of the simulation, the relative energy density in gravitational radiation is fairly steady at a value of $\rho_\mathrm{GW}\simeq 2\times 10^{-4}\bar\rho_\mathrm{DM}$.

If continued PBH mergers were to maintain the same relative level of gravitational radiation up to $z=0$, the energy density parameter of the gravitational radiation would be $\Omega_\mathrm{GW}\simeq 5\times 10^{-5}$ today.
Alternatively, neglecting all gravitational-wave production after the end of the simulation leads to $\Omega_\mathrm{GW}\simeq 10^{-5}$ today.
LVK limits on a gravitational-wave background \cite{KAGRA:2021kbb} are frequency-dependent but generally lie around $\Omega_\mathrm{GW}\lesssim 10^{-8}$ per e-fold in frequency.
Thus, even if only a small fraction of the gravitational radiation were to lie in the LVK frequency range, the PBH scenario that we consider would be ruled out.

This conclusion assumes that the simulated region is typical. As we noted in section~\ref{sec:simulations}, our simulation region is somewhat special because it is centered on a $3\sigma$ density excess in $10^4$ PBHs (about $10^5~\Msol$).
More importantly, the simulation volume is small enough that it excludes most of the standard adiabatic contributions to the matter power spectrum (see figure~\ref{fig:linearpower}).
Thus, we effectively study a rare region in which the adiabatic modes are small.
A larger simulation volume is needed to assess the impact of the large-scale adiabatic structure on the PBH dynamics that lead to gravitational radiation.
For example, halos of much larger mass than those in our simulation would begin to form around $z\sim 30$, when the adiabatic modes start to become nonlinear.

However, we can consider a very conservative assumption that the production of gravitational waves stops at $z=30$. As we show in figure~\ref{fig:GW}, this assumption leads to $\Omega_\mathrm{GW}\sim 10^{-6}$ today.
This energy density parameter still vastly exceeds the LVK limit -- but only if enough of the radiation produced at these redshifts lies within the appropriate frequency range.
In a followup paper, we will characterize more precisely the gravitational radiation produced by the PBH mergers.

\section{Summary}\label{sec:summary}

We have presented the first simulation of cosmological structure formation out of PBH dark matter that consistently incorporates few-body dynamics as well as relativistic effects. The simulation volume is of comoving size $\sim 40\,{\rm kpc}$ and contains of order 100,000 PBHs of mean mass $16.5\, \Msol$. We have carefully constructed the initial conditions, including initializing the simulations with a population of early-formed binaries.
Our main findings can be summarized as follows.

The random initial spatial distribution of the PBHs gives rise to a white-noise (Poisson) isocurvature contribution to the power spectrum, leading to significant halo formation by $z\sim  400$. However, we find that collisional dynamics suppress the abundance of these halos.
Collisional dynamics also decrease the internal density of PBH halos over time, in contrast to a collisionless case where the high initial density of early-forming halos is roughly preserved.
Inside a PBH halo, collisional relaxation causes mass segregation on a time scale of order the two-body relaxation time, and central cores form and grow over a time scale about an order of magnitude longer than the relaxation time.

We also identify a backreaction from small-scale PBH dynamics onto cosmic structure at much larger scales.
High-speed ejections of PBHs from many-body interactions create a subcomponent of hot dark matter. For our scenario with $\mathcal{O}(10~\Msol)$ PBHs, we expect this collisional heating to have a non-negligible effect on structure formation up to mass scales exceeding $10^{10}~\Msol$. 
Because such large scales are not represented in our simulation, further study is needed to characterize this effect precisely.
However, this could provide an additional, purely gravitational channel to constrain the PBH paradigm.

Our simulation resolves binary coalescence events, and we find that analytic predictions for the timings of these events correlate extremely poorly with the actual merger times.
Although the initial orbital parameters of binary pairs are accurately predicted by the analytic model of ref.~\cite{Raidal:2018bbj}, these parameters change completely as time goes on due to interactions with other PBHs.
This outcome implies that the PBH merger rate, and the resulting gravitational-wave signal, are unlikely to be accurately predicted using prior analytic models.
Also, we find that only two second-generation mergers occurred during the simulation, which suggests that models including cascades of PBH mergers may be unlikely, at least for a Poissonian initial PBH distribution.

Finally, we calculate the total energy in gravitational waves produced during the simulation and find that, even under quite conservative assumptions, this background is likely to be larger than current experimental (LVK) bounds.
However, for precise conclusions, one needs to consider the frequency spectrum of the emitted gravitational radiation and its overlap with the LVK band. We leave a detailed investigation to upcoming work.

\acknowledgments

The authors thank Thorsten Naab for discussions during the early stages of this project, Carl Rodriguez for useful discussions on collisional dynamics, and Gabriele Franciolini and Karsten Jedamzik for helpful comments on the manuscript. Simulations for this work were carried out on the \textsc{Freya} cluster hosted by the Max Planck Computing and Data Facility in Garching, Germany. SY is supported by STFC grant ST/X000796/1.

\appendix

\section{Simulations with radiation and baryon backgrounds}\label{sec:simulation-backgrounds}

This appendix details how we model the evolution of the PBH distribution within a background of radiation and baryonic matter. As discussed in section~\ref{sec:collisionless_simulations}, we assume that both of these components are spatially uniform on the scales that we study.

\subsection{Perturbation theory}\label{sec:linear}

We first point out how subhorizon density perturbations evolve at linear order.
Here we consider dark matter density perturbations within a homogeneous background of radiation and baryons. Let $\aeq\simeq 2.94\times 10^{-4}$ be the scale factor of matter-radiation equality, and let $\fb\simeq 0.157$ be the fraction of the matter that consists of baryons, so that $1-\fb$ is the dark matter fraction. The general solutions to the equations governing the evolution of a dark matter density perturbation $\delta$ are then \cite{Hu:1995en}
\begin{align}\label{Dgen}
    D_\pm(a)=(1+a/\aeq)^{\mu_\pm}\,_2F_1\!\left(-\mu_\pm,\frac{1}{2}-\mu_\pm,\frac{1}{2}-2\mu_\pm,\frac{1}{1+a/\aeq}\right),
\end{align}
where $_2F_1$ is the hypergeometric function and
\begin{equation}
    \mu_\pm\equiv\pm\frac{5}{4}\sqrt{1-\frac{24}{25}\fb}-\frac{1}{4}.
\end{equation}
Note that $\mu_+\simeq 0.901$ and $\mu_-\simeq -1.401$.
Each perturbation evolves proportionally to some particular combination
\begin{align}\label{Dsum}
    D(a) = A_+ D_+(a) + A_- D_-(a).
\end{align}

\paragraph{Isocurvature modes}
The initial Poisson clustering of PBHs manifests as ``isocurvature'' modes, which are initially constant during radiation domination. This initial condition demands that
\begin{align}\label{Diso}
    A_\pm = 
    \frac{-4^{\mu_\pm}\sqrt{\pi}\,\Gamma(-2\mu_\pm)}
    {\Gamma(1/2-2\mu_\pm)\left[\psi(-2\mu_\pm)-\psi(-2\mu_\mp)\right]}
\end{align}
for these modes, where $\Gamma(x)$ is the gamma function and $\psi(x)\equiv\diff\ln\Gamma(x)/\diff x$ is the digamma function. For convenience, we have normalized these coefficients so that $D(a)=1$ for $a\ll\aeq$.
It is convenient to note that
\begin{align}\label{isogrowth_MD}
    D(a)\to A_+(a/\aeq)^{\mu_+} \simeq 2094 a^{0.901}
\end{align}
deep in matter domination (taking $a=1$ today).

\paragraph{Adiabatic modes}
There are also ``adiabatic'' modes sourced by the primordial power spectrum, which we take to be statistically uncorrelated with the isocurvature modes. Adiabatic modes arise because the radiation is initially inhomogeneous.
During horizon entry, it imparts a gravitational kick on the dark matter. Although the peculiar gravitational potentials sourced by the radiation decay rapidly after horizon entry, the dark matter remains in motion due to the initial kick, and density perturbations in the dark matter slowly grow in amplitude. For these perturbations, ref.~\cite{Hu:1995en} showed that
\begin{align}\label{Dadi}
    A_\pm = 
    \frac{
    -\Gamma(-\mu_\pm)\Gamma(1/2-\mu_\pm)
    \left[B(k)+ 2\psi(1)-\psi(-\mu_\mp)-\psi(1/2-\mu_\mp)\right]
    }
    {\Gamma(1/2-2\mu_\pm)(\psi(-\mu_\pm)+\psi(1/2-\mu_\pm)-\psi(-\mu_\mp)-\psi(1/2-\mu_\mp))},
\end{align}
where sensitivity to the perturbation wavenumber $k$ is contained in the function
\begin{align}
    B(k) \equiv \ln\!\left[0.594 \left(1 - 0.631 \fnu + 0.284 \fnu^2\right)\frac{4(k/\keq)^2}{1 + \sqrt{1 + 8(k/\keq)^2}}\right]
\end{align}
and arises because different modes enter the horizon at different times.
Here, $\keq=\aeq H(\aeq)$ is the wavenumber entering the horizon at $\aeq$ (where $H(a)$ is the Hubble rate at $a$) and $\fnu\simeq 0.409$ is the fraction of the radiation that consists of neutrinos.

\subsection{Simulations in comoving coordinates}

We carry out collisionless simulations in comoving coordinates using the \textsc{Gadget}-4 simulation code.
Homogeneous baryon and radiation components with density parameters $\OmegaB\simeq 0.0490$ and $\OmegaR\simeq 9.14\times 10^{-5}$, respectively, are straightforwardly modeled in this context by simply updating the code to evaluate the Hubble rate as
\begin{align}
    H = H_0\sqrt{(\OmegaC+\OmegaB)a^{-3}+\OmegaR a^{-4}+\OmegaL}.
\end{align}
Here $\OmegaC\simeq 0.262$ is the usual density parameter of the simulation particles, which represent dark matter, and $\OmegaL\simeq 0.689$ is the density parameter of dark energy. For simplicity, we neglect variation in the number of relativistic degrees of freedom during the radiation epoch.

We generate adiabatic initial perturbations using the Zel'dovich approximation. Given a density field $\delta(\vec k)$ in Fourier space, the field of comoving particle displacements is $\vec s(\vec k)=(\I\vec k/k^2)\delta(\vec k)$ as usual. However, initial comoving velocities are now
\begin{align}
    \dot{\vec{s}}(\vec k) = H\frac{\diff\ln D}{\diff\ln a}\vec s(\vec k),
\end{align}
where $D$ is the ($k$-dependent) adiabatic growth function given above
(see appendix~A of ref.~\cite{Delos:2018ueo} for further detail).

To validate this approach, we carried out simulations of the periodic 101-kpc particle dark matter and PBH boxes described in section~\ref{sec:collisionless_ics}, starting from the initial scale factor $a\simeq 2.9\times 10^{-12}$. Although both boxes are only explicitly initialized with adiabatic perturbations, the PBH box naturally also includes large-amplitude isocurvature perturbations due to the independently randomly distributed particle positions and masses. In fact, the initial power spectrum of the isocurvature modes is
\begin{equation}
    P(k)=\frac{1}{\bar\rho_\mathrm{DM}^2}\int_0^\infty\diff m\frac{\diff n_\mathrm{PBH}}{\diff m}m^2,
\end{equation}
where $\bar\rho_\mathrm{DM}$ is the mean density of the dark matter and $\diff n_\mathrm{PBH}/\diff m$ is the mass function of the PBH particles, i.e., their differential number density per mass interval and spatial volume.

\begin{figure}
  \includegraphics[width=\textwidth]{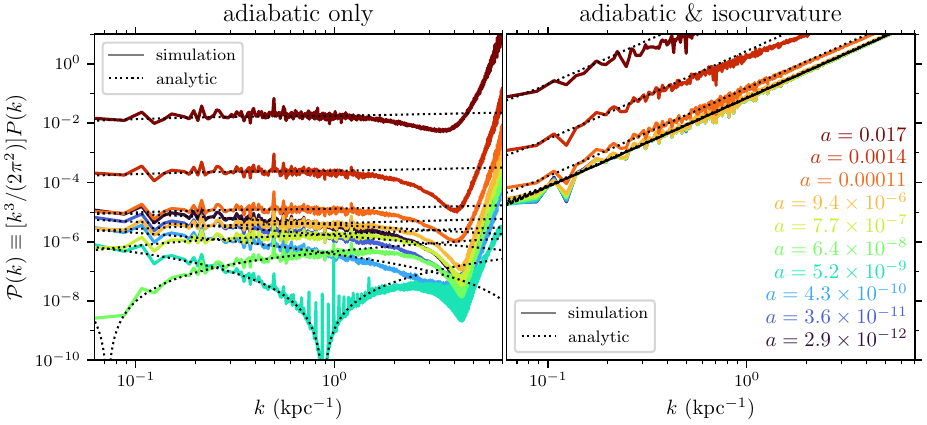}
\caption{Comparing the evolution of the dark matter density power spectrum in simulations (solid curves) to analytic predictions (dotted curves). We consider a range of times (different colors). For the left-hand panel, we use the simulation of an adiabatically perturbed initial grid, representing particle dark matter. Since this simulation only includes adiabatic perturbations, we compare it to evolution according to the adiabatic growth function in appendix~\ref{sec:linear}. For the right-hand panel, we use the simulation of an adiabatically perturbed PBH distribution, which includes both adiabatic and Poissonian isocurvature perturbations, so we compare it to the appropriate combination of adiabatic and isocurvature growth functions. The simulations tightly match analytic predictions, except when $\mathcal P\gtrsim 1$, due to nonlinear effects, and at high $k$ in the left-hand panel, due to grid artifacts.}
\label{fig:power}
\end{figure}

\begin{figure}
  \includegraphics[width=\textwidth]{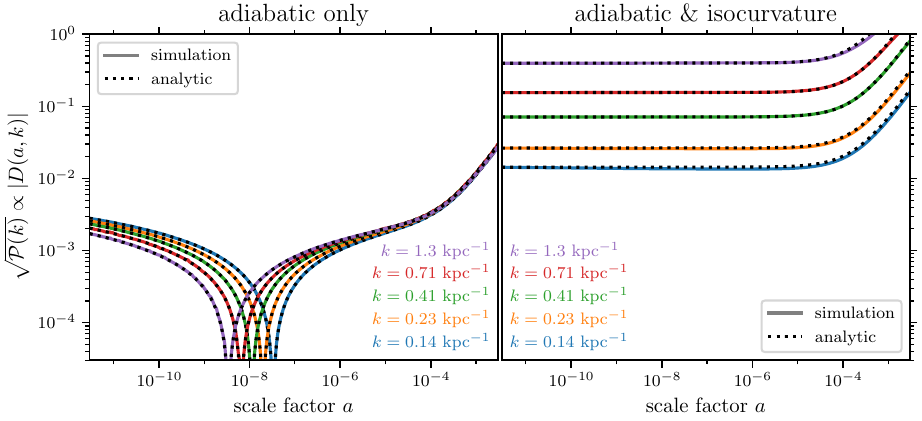}
\caption{Similar to figure~\ref{fig:power} but comparing the time evolution of $\sqrt{\mathcal P(k)}$ (solid curves) to that predicted by the analytic growth functions (dotted curves) for several $k$ (different colors). Evolution in the simulations agrees well with analytic predictions.}
\label{fig:growth}
\end{figure}

Figure~\ref{fig:power} shows the evolution of the dark matter power spectra in the particle dark matter and PBH simulations, specifically its dimensionless form $\mathcal{P}(k)\equiv [k^3/(2\pi^2)]P(k)$.
We also show analytic predictions for comparison, which are evolved according to the appropriate linear growth functions (appendix~\ref{sec:linear}). For the particle dark matter simulation, we compare the linear-theory adiabatic power spectrum only, while for the PBH simulation, we compare the sum of adiabatic and isocurvature power spectra, since the adiabatic and isocurvature perturbations are statistically uncorrelated. However, the isocurvature power totally dominates over the adiabatic power.
The match between simulations and linear-theory predictions is evidently very strong except at high $k$ for the particle dark matter simulation, where artifacts of the initial grid become important, and when $\mathcal{P}\gtrsim 1$, when density perturbations are becoming nonlinear.
Figure~\ref{fig:growth} shows a similar test, in this case comparing for several wavenumbers $k$ the evolution of $\sqrt{\mathcal{P}(k)}$ to the appropriate analytic growth functions $|D(a)|$. Here, again, there is a tight match between the simulations and the analytic predictions as long as $\sqrt{\mathcal{P}}\lesssim 1$.

Note that the adiabatic growth functions $D(a)$ cross zero and change sign at sufficiently early times. This effect is visible in figures \ref{fig:power} and~\ref{fig:growth} as a dip toward zero for the particle dark matter simulations, since these figures show $|D|$ (or its square). It is not physically correct, because it occurs before horizon entry, when both relativistic effects and radiation perturbations (both neglected in the growth functions given in equation~\ref{Dgen}) are important.
Conceptually, this sign switch occurs because we are extrapolating into the past by assuming that particles drift freely even at times before they were given their initial kicks. Thus, the particles pass through their initial (Lagrangian) positions, leading to a sign flip in the density contrasts.
Although this effect does not really occur in the Universe, its presence is not a cause for concern. As figures \ref{fig:power} and~\ref{fig:growth} show, the correct evolution is reproduced even for modes for which this sign flip occurred.

\subsection{Simulations in physical coordinates} \label{section:physical-coordinates}

In physical coordinates at subhorizon scales, a homogeneous background of density $\rho$ and pressure $P$ drives a gravitational acceleration
\begin{equation}\label{key}
	\vec f = -\frac{4}{3}\pi G (\rho+3P)\vec r
\end{equation}
at the position $\vec r$. For radiation, baryons, and dark energy with density parameters $\OmegaR$, $\OmegaB$, and $\OmegaL$, respectively, this acceleration is
\begin{align}
	\vec f &= -\frac{4}{3}\pi G\rho_\mathrm{crit}\left(\OmegaB a^{-3} + 2\OmegaR a^{-4} - 2\OmegaL \right)\vec r
	\\\label{background_force}
	&= -\frac{H_0^2}{2}\left(\OmegaB a^{-3} + 2\OmegaR a^{-4} - 2\OmegaL \right)\vec r
\end{align}
as a function of the scale factor $a$. By integrating $\diff t=\diff a/(aH)$ to obtain $a$ as a function of time, we implement equation~(\ref{background_force}) as a time-dependent external force within the collisional PBH simulation.
The accuracy of this implementation is confirmed by the tight match between the collisional and collisionless simulations at early times, before nonlinear structure starts to dominate.

\section{Binary systems in the collisional initial conditions}\label{sec:simulation-binaries}

\begin{figure}
  \centering
  \includegraphics[width=.98\textwidth]{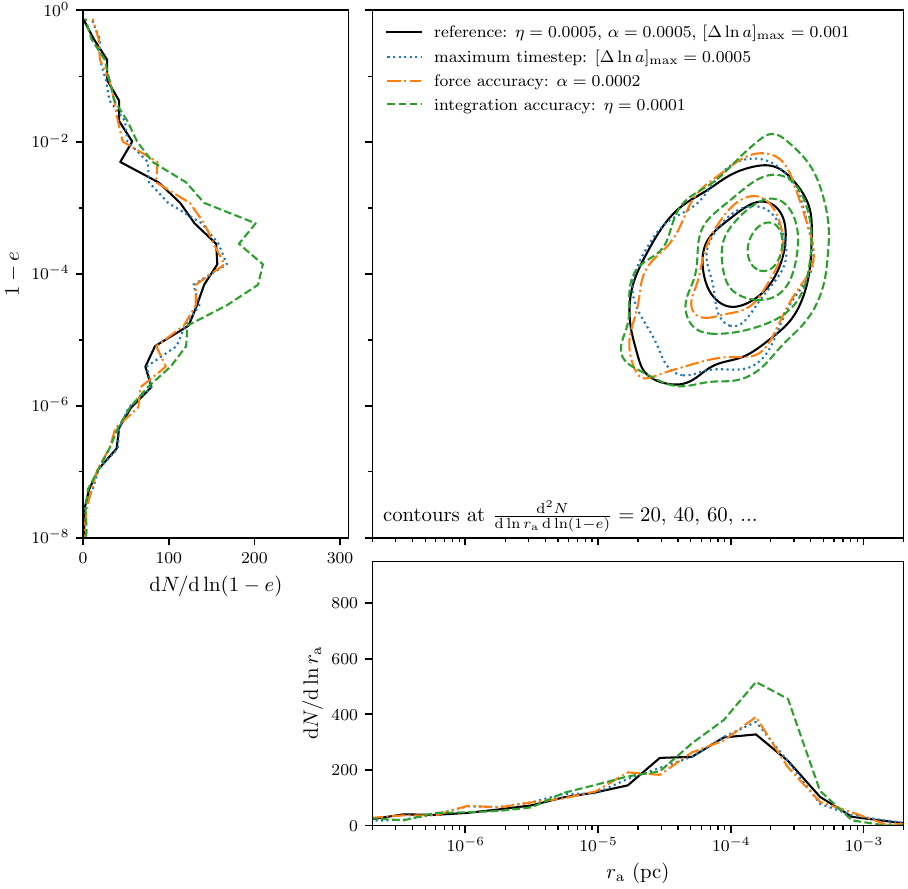}
\caption{
    Convergence tests of the distribution of PBH binaries identified by $a=10^{-5}$. The main panel shows the differential number of binary systems, $\frac{\diff^2 N}{\diff\ln r_\mathrm{a}\diff\ln(1-e)}$, per interval in eccentricity $e$ and semi-major axis $r_\mathrm{a}$.
    The other panels show the projected distributions for each of these parameters.
    For these parameters, the binary distribution is converged well with respect to the force accuracy parameter and the maximum time step, but it is not converged with respect to the integration accuracy parameter $\eta$.
    }
\label{fig:binaries_a}
\end{figure}

\begin{figure}
  \centering
  \includegraphics[width=.98\textwidth]{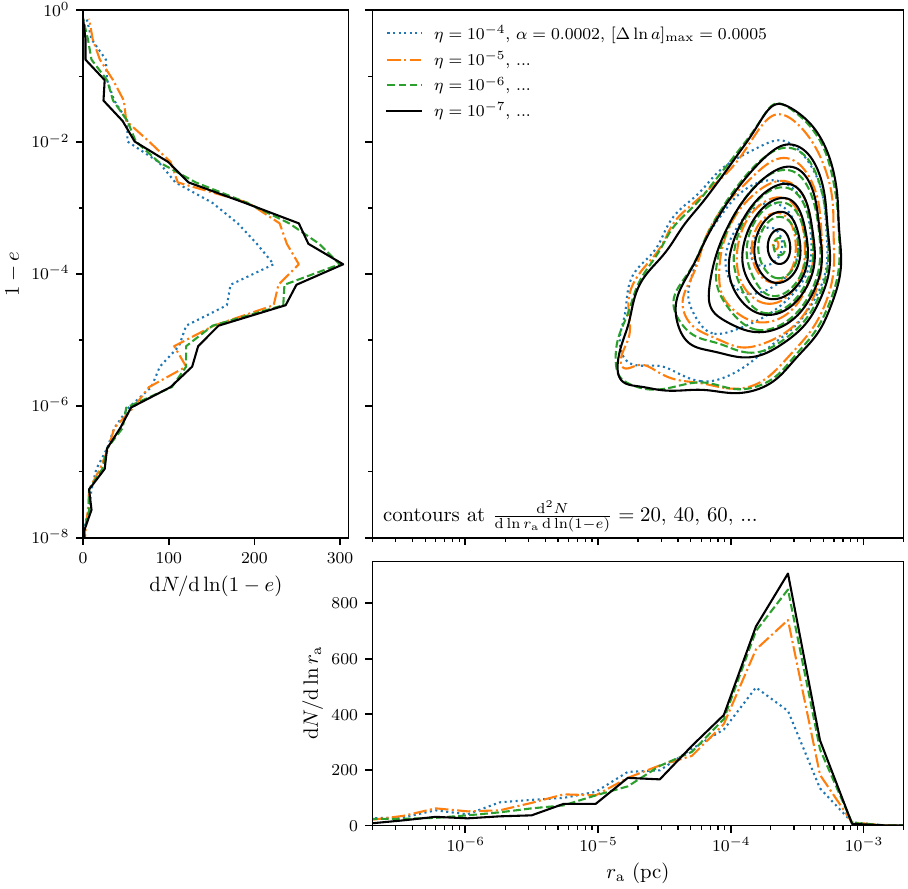}
\caption{
    Similar to figure~\ref{fig:binaries_a} but demonstrating convergence with respect to the integration accuracy parameter $\eta$ for $\eta=10^{-7}$.
    }
\label{fig:binaries_b}
\end{figure}

From the initial time, $a\simeq 3\times 10^{-12}$, up to $a=10^{-5}$, the collisional PBH simulation is executed using \textsc{Gadget-4}, the same cosmological simulation code used for the collisionless simulations. For this purpose, the following numerical parameters are employed.
\begin{itemize}
    \item The force accuracy parameter is $\alpha=0.0002$ (compared to $\alpha=0.005$ in the collisionless simulations). This parameter represents approximately the maximum allowed fractional error in the tree-based force evaluation (see equation~36 of ref.~\cite{Springel:2020plp}).
    \item The integration accuracy parameter is $\eta=10^{-7}$ (compared to $\eta=0.025$ in the collisionless simulations). The meaning of this parameter is roughly that the displacement induced by gravitational acceleration over a time step is smaller than $\eta$ times the force softening length (see equation~81 of ref.~\cite{Springel:2020plp}).
    \item The maximum time step is $\Delta\ln a=0.0005$ (compared to $0.01$ in the collisionless simulations).
    \item The comoving softening length is 0.6~pc (compared to 24~pc in the collisionless simulations), so forces are sub-Newtonian at comoving distances smaller than 1.7~pc (67~pc in the collisionless simulations).
\end{itemize}
Despite these tightened accuracy parameters, the code still cannot be expected to resolve the dynamics of the many PBH binaries that are expected to form by $a=10^{-5}$.
Therefore, whenever a pair of PBHs lies within 3.4~pc during a time step (twice the distance at which forces become sub-Newtonian), we record their positions and velocities and determine the associated Keplerian orbit. At $a=10^{-5}$, when we switch to the \bifrost{}\ collisional simulation code, we modify each of these pairs to move it into the appropriate phase of this Keplerian orbit, while retaining the center-of-mass position and velocity of the pair. For this procedure, we neglect binary pairs for which the Keplerian orbit is unbound (hyperbolic), but there are only three such pairs identified.

\begin{figure}
  \centering
  \includegraphics[width=.6\textwidth]{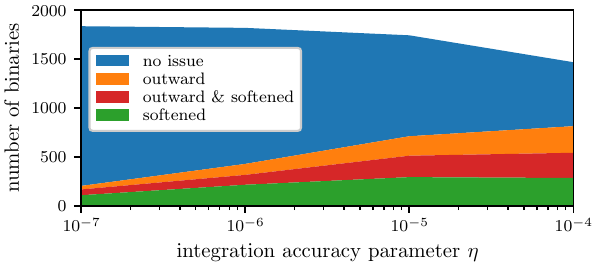}
\caption{Number of PBH binaries identified by $a=10^{-5}$ as a function of the integration accuracy parameter $\eta$. The total count is converged for $\eta=10^{-7}$ to a value of 1837. 204 of them were identified while inside the distance at which forces are sub-Newtonian (green), already moving away from each other (orange), or both (red). The properties of these subsets of the binary systems are likely to be inaccurate.}
\label{fig:binaries_eta}
\end{figure}

In figures \ref{fig:binaries_a} and~\ref{fig:binaries_b}, we explore how the distribution of these binary orbits depends on the numerical parameters of the simulation.
Figure~\ref{fig:binaries_a} shows that for the force accuracy parameter $\alpha=0.0002$, maximum time step $\Delta\ln a=0.0005$, and integration accuracy parameter $\eta=10^{-4}$, the distribution of binaries is converged with respect to $\alpha$ and the maximum $\Delta\ln a$, in the sense that the binary distribution is the same when the respective parameters are set to these values as when they are set to larger (less stringent) values. However, the distribution is not converged with respect to $\eta$.

Figure~\ref{fig:binaries_b} shows that, with respect to the integration accuracy parameter $\eta$, a reasonable degree of convergence is achieved once $\eta=10^{-7}$. A small fraction of the binary systems still have unreliable properties, however.
In particular, a binary system's properties are definitely incorrect if the binary pair is first identified, and its Keplerian orbit recorded, after the pair already lies within the distance at which forces are sub-Newtonian (due to softening). Additionally, if a pair is identified while the two PBHs are moving apart (instead of together), this means that their pericenter was completely missed in the time integration, and the binary orbit is likely to be inaccurate.
As a function of $\eta$, figure~\ref{fig:binaries_eta} shows the total count of binaries identified by $a=10^{-5}$ along with how many are unreliable in the aforementioned ways. The total number of binaries is converged well by $\eta=10^{-7}$ at a value of 1837, but a little over 10 percent of these -- 204 -- still have unreliable properties.
When we analyze the properties of early binary systems in section~\ref{sec:binaries}, we omit these unreliable binaries.

\bibliographystyle{JHEP}
\bibliography{main}

\end{document}